\documentclass{article}

\usepackage{cite}
\usepackage{arxiv}
\usepackage{slashed}
\usepackage[utf8]{inputenc} 
\usepackage{fontenc,amssymb }
\usepackage{hyperref}       
\usepackage{url}            
\usepackage{booktabs}       
\usepackage{amsfonts}       
\usepackage{nicefrac}       
\usepackage{microtype}      
\usepackage{lipsum}
\usepackage{physics}
\usepackage{graphicx}
\usepackage{mathtools}
\title{Accelerated paths and Unruh effect I:  scalars and fermions in Anti De Sitter spacetime }

\author{
 Shahnewaz Ahmed{} \\
  School of Data and Sciences\\ 
  BRAC University,  \\
  66 Mohakhali, Dhaka 1212. Bangladesh\\
  {\tt shahnewaz.ahmed@bracu.ac.bd}\\
  {\tt ahmed.shahnewaz16@gmail.com}\\
   \And
 Mir Mehedi Faruk\\
 Department of Physics\\
  McGill University,\\
Montreal, QC H3A 2T8, Canada\\
  \texttt{mir.faruk@mail.mcgill.ca} \\}
\begin{document}
\maketitle
\begin{abstract}
We have investigated  the Unruh effect in Anti de-Sitter (AdS) spacetime by
examining the response function
 of an    Unruh-DeWitt particle detector
 with uniform constant  acceleration. 
An exact expression of the detector response function for the scalar field
has been obtained
with different levels of non-linearity in even dimensional AdS spacetime. We also showed how the response of the accelerated Unruh detector  coupled quadratically to massless Dirac field in $D$ dimensional $(D \geq 2)$ AdS spacetime is proportional to that of a detector linearly coupled to a massless scalar field in $2D$ dimensional AdS spacetime. Here, the fermionic and scalar matter field is coupled minimally and conformally to the background AdS metric, respectively. Finally, we discuss about 
the extension of the 
results for more general 
stationary motion.\\\\
\end{abstract}


The Unruh effect describes a phenomenon where uniformly accelerated observers with constant proper acceleration observe a thermal bath of particles in the no 
particle/vacuum state of inertial observers\cite{review 1, review 2, Weinberg72}.
Although this phenomenon was initially discovered for flat spacetime, this has been extended to spacetimes with curvature and cosmological horizon\cite{review 1,GH} as well.
In curved spacetime,
interesting relations are noted  between Unruh temperature and cosmological constant $\Lambda$.
For example,
in de Sitter (dS) spacetime,
a comoving detector
 has a thermal response at a temperature
 referred as the Gibbons-Hawking temperature,
 $T=\frac{\sqrt{\Lambda/3}}{2\pi}$
 \cite{GH, DSNEW, Edu1, Edu2}. Following these developments,  Deser 
 and Levin\cite{D&L}
  explained further the Unruh effects in
  maximally symmetric curved spacetimes, i.e. de Sitter and Anti de Sitter (AdS) space time
 using the Global Embedding Minkowski Spacetimes
 (GEMS) approach\cite{GEMS, GEMS2, GEMS3}
 and their relations to 
 black hole physics\cite{Padda2}.
 Now the special feature in Minkowski space is that it has a unique vacuum. This is not, in general, a property of 
 curved spacetimes.
 In order to understand the Unruh effect better in curved spacetime Unruh and Dewitt\cite{BD,UD1,UD2} constructed a concept of particle detector in order to examine the Unruh radiation\footnote{Historically there were other constructions before it, for example reference \cite{BD}. See chapter 3 of ref. \cite{BD} and \cite{Weinberg72} for more a detailed description of particle detectors.}.  The early\cite{early} and late\cite{late} time physics of quantum fields in curved spacetimes with a cosmological constant has been studied extensively.
 Spacetimes with positive cosmological constant has direct application in cosmology\cite{mukha} and particle physics\cite{linde}
unlike the spacetimes with negative cosmological constants such as Anti De Sitter (AdS) spacetime. However, quantum gravity theories in AdS background  are  studied
in great extent due to the discovery of  AdS/CFT correspondence\cite{Witten}.\\ \\
AdS spacetime is a maximally symmetric spacetime with negative constant scalar curvature
which
provide a unique prescription for understanding quantum gravity theories from a holographic perspective. 
Holographic techniques have been immensely successful for studying strongly coupled quantum field theories\cite{quark dynamics,adscft}. More precisely the AdS/CFT correspondence
dictates that the  theory of
quantum gravity in asymptotically $AdS_{D+1}$ spacetime is dual to an ordinary $CFT_D$ without
gravity.
These dualities between strong and weak coupling play an important role in the current
paradigm of string and quantum field theories, although explicit demonstrations of such dualities exist
only for a limited number of cases. One  precise explicit  example of such holographic principle  is 
the duality between type $IIB$ string  theory on $AdS_5\times S_5$ and   $\mathcal{N}=4$ super Yang-Mills in four dimensions. Using such strong/weak dualities, strongly coupled quantum field theories are much better understood.
The study of the Unruh effect in holographic duality has been discussed in different contexts including  quark dynamics\cite{spinorExchange}, Brownian motion\cite{brownian motion}, entanglement entropy\cite{EE}, black holes\cite{unruhbh}, holographic complexity\cite{complexity}, Casimir effect\cite{saha} etc.
Therefore it is important to understand the features of quantum field theory and thermality relations in AdS spacetimes\cite{bros}.\\\\There has been several efforts to understand the detector response of 
constantly 
accelerating detectors in several contexts\cite{new,sri, Ted Jacobson, last}.
 Explicit computations are available for the detector response function in $D$ dimensional Minkowski spacetime where a peculiar trend of ''inversion of statistics" has been noticed in odd dimensions for linearly coupled detectors with a scalar field\cite{takani}.
 This means the detector response function for scalars due to uniform acceleration 
takes the form of Fermi-Dirac distribution
instead of a Bose-Einstein distribution.
Following Sewell's result\cite{Sewell}, Takagi\cite{takagiReview} has shown this
kind of 
behaviour 
for a free field and Ooguri\cite{Ooguri} concluded the arguments for the interacting theory. In the case of Dirac fields in Minkowski spacetime, Louko and Toussaint \cite{Nottingham} proved that uniformly accelerated detectors detect Bose-Einstein distributions in any dimension $D$. Based on the Unruh effect observed by accelerated detectors, several construction of quantum heat engine\cite{ottoEngine,engine2} have been proposed for flat space.\\\\Calculating the response function in AdS spacetime  is much harder compared to flat spacetimes. 
Jennings\cite{Jennings}
 analyzed how the spacetime AdS curvature and dimensionality affect the response spectrum of an accelerated detector linearly coupled to 
a scalar field.
Indeed,
it has already been shown in ref.
\cite{Ted Jacobson, last, Jennings} that in AdS spacetime thermality arises in the free field case when  acceleration is above the mass scale of the 
AdS curvature.
 We should note here is that AdS spacetime has a boundary. In Poincare coordinate (\ref{metric}) this boundary corresponds to
  $z\rightarrow0$ limit. So we have considered all the constant accelerated paths in AdS spacetime which is confined in a two dimensional plane inside a higher dimensional embedding space\cite{bros}. 
   Jennings\cite{Jennings} has studied  features
of   inversion of statistics
   of scalar fields
linearly
coupled to uniformly accelerated 
detectors 
in the $z-t$ plane of
AdS spacetime.
In this article, we first extend the results of ref. \cite{Jennings}
for scalar fields by considering all possible linear constant (uniform) acceleration. We have  also considered more general non-linear coupling between the scalar field and the detectors \eqref{kukumia}.
Such non linear coupling has been studied in flat spacetime\cite{sri,ottoEngine,engine2}
but not carefully studied in curved spacetime such as AdS.
 We examine the effect of nonlinear coupling
on the KMS condition, thermality and statistics inversion.
In section 1.1 we elaborate on how to compute the response function for uniform acceleration in even dimensional AdS spacetime.
We derive analytic expression for the response function for any level of non-linearity.
We concentrate on the fermions in the second section.
Some interesting conclusion has been drawn in ref. \cite{Nottingham}
who
showed that the response 
of uniformly accelerating Unruh detector coupled quadrat
ically to Dirac fields in $D$ dimensional Minkowski spacetime
is identical to that
of a detector coupled linearly and conformally to a massless scalar field in $2D$  dimensional Minkowski spacetime.
We investigate such claim for Dirac fields in AdS spacetime
(see theorem 2).
Furthermore, we have discussed about the   possibility to extend the 
results 
for more general stationary trajectories (see Appendix C for detailed analysis).
The response function  of scalar fields in four dimensions for different values of $n$ has been analyzed in detail (figure 2-4). At the end we compare the fermionic response function with the bosonic one (figure 5). 
Finally we conclude the article with some remarks and future prospects we would like to pursue.
\section{
 Scalar field in AdS}
\label{sec:intro}
We first study a real scalar field $\Phi$ in $D$ dimensional AdS spacetime
which is conformally coupled to gravitational background. Here we
are considering the AdS metric in Poincare coordinates but 
we can take any other coordinates system such as global coordinates. We employed $\hbar = 1$, $c=1$ and Boltzmann constant $k_B = 1$ in our calculation. 
The AdS metric in Poincare coordinate,
\begin{eqnarray}
ds^{2}=\displaystyle \frac{1}{k^{2}z^{2}} (dt^{2}-dx_{1}^{2}-dx_{2}^{2} - ... -dx_{D-2}^{2} - dz^2).\label{metric}
\end{eqnarray}
Here $k$ is the curvature of the spacetime and related to the negative cosmological constant through $|\Lambda|=k^{2}(D-1)(D-3)/2$ and Ricci scalar $R = -D(D-1)k^2$.
Alternatively the metric can also be wriiten as,
\begin{equation}
ds^{2}=e^{-2yk}(
dt^{2}-dx_{1}^{2}-dx_{2}^{2} - ... -dx_{D-2}^{2}
)-dy^{2}.
\label{ds2deSit}
\end{equation}

The action we are interested is-
\begin{eqnarray}
S_0=\frac{1}{2}\int d^Dx \sqrt{|g|}\left(g^{\mu\nu}\triangledown_\mu \Phi
\triangledown_\nu \Phi+\zeta R\Phi^2\right).
\end{eqnarray}
For conformally coupled
scalars to gravity we can choose\cite{Weinberg72},
\begin{eqnarray}
\zeta= \frac{D-2}{4(D-1)}. \label{zeta}
\end{eqnarray}
The corresponding eq of motion is,
\begin{eqnarray}
\left(\nabla_\mu \nabla^\mu +\frac{D-2}{4(D-1)} R \right)\Phi(x)=0 \label{scalareqn}
\end{eqnarray}
We can identify the Klein Gordon wave operator from the above equation,
\begin{eqnarray}
\Box^{^{KG}}_x =   \bigg( g^{\mu \nu} \partial_{\mu} \partial_{\nu} - \frac{D-2}{z}  \partial^{z} - k^2\frac{D(D-2)}{2} \bigg).
\end{eqnarray}
Consider the detector coupled  to a real
scalar field $\Phi$,  through the  following interaction Lagrangian. Also, the total action of the  system is\footnote{Readers can go through ref. \cite{Jennings} for further description of $S_{detector}$.},\\
\begin{eqnarray}
&&\mathcal{L}_{int}=c m(\tau)\Phi^n[x(\tau)] \label{kukumia}\\
&&S=S_0+S_{int}+S_{detector},
\end{eqnarray}
where, $n$ is any positive integer. $m(\tau)$ is the detector’s monopole moment and $c$ is a small coupling constant. Choosing $n=1$ gives us usual linearly coupled Unruh-DeWitt detector\cite{Jennings}.
The two point correlators can easily be  obtained
in the following form with suitable boundary condition\cite{Jennings},
\begin{eqnarray}
G_{\rm AdS_{D}}(x,x')=
\bra{0}\Phi(x(\tau))
\Phi(x(\tau'))\ket{0}={\cal C}_{D} \bigg(\frac{1}{(v-1)^{D/2-1}}-\frac{1}{(v+1)^{D/2-1}}\bigg)\label{two}
\end{eqnarray}
where, 
\begin{eqnarray}
&&{\cal C}_{D}= \frac{k^{D-2}\Gamma(D/2-1)}{2(2\pi)^{D/2}}, \label{CDCD}\\
 && v = \frac{z^{2}+z^{\prime 2}+(\mathbf{x}-\mathbf{x}')^{2}-(t-t'-i\epsilon)^{2}}{2zz'}\label{nu}.
\end{eqnarray}
The detector moves along the worldline $x(\tau)$ in the $D$-dimensional AdS spacetime with constant acceleration. For our current discussion we are taking uniform accelerated paths. In AdS spacetime any timelike path with constant acceleration $a$ can be categorized in three ways\cite{Jennings} depending upon AdS curvature
$k$-\\ \\
(i) sub critical paths $(a<k)$,\\
(ii) critical paths $(a=k)$,\\
(iii) super critical paths $(a>k)$.\\ \\
We are  considering the supercritical paths as only these paths results in non zero response function for the detectors\cite{Jennings} in uniform linear motion.
Using global embedding approach, 
one can construct a path
with constant acceleration
by considering it as a intersection between a $M(M<D+1)$ dimensional flat plane and $D$ dimensional AdS
hypersurface embedded in $D+1$ dimensional flat spacetime.
To obtain linear uniform supercritical motion one have to work with the conic section constructed using $M=2$ dimensional plane and AdS hypersurface. In appendix C we have employed global embedding approach to show that for all uniform linear supercritical trajectories confined between the intersection of a two dimensional plane and AdS hypersurface would have the same conformal invariant $v$ as a function of proper time.  
\begin{eqnarray}
v(\tau,\tau')=\frac{a^2}{\omega^2}- \frac{k^2}{\omega^2}\cosh(\omega(\tau-\tau')-i\epsilon)\label{path2}.
\end{eqnarray} 
In equation (\ref{path2}) we introduced $i\epsilon$ prescription. We demonstrate  example of two such trajectories in local Poincare coordinates. We will refer them as timelike particle trajectories in Poincare coordinates along the $x^1-t$ plane and the $z-t$ plane. 
The example of super critical
path (with constant linear acceleration) in  $z-t$ plane is given in ref. \cite{Jennings},
\begin{equation}
t(\tau)=\frac{a}{\omega}e^{\omega \tau} \;\;,\;\;
z(\tau)= e^{\omega\tau} \;\;,\;\;
x^{1} =x^{2}=x^{3}=\ldots=x^{D-2}=0. \label{eq12}
\end{equation}
We can also define a path like this in the $x_1-t$ plane\footnote{See the appendix to find out such path also results in  constant acceleration.},
\begin{eqnarray}
z(\tau)=z_0\;\;,\;\;
x^{1}(\tau)=\frac{z_0k}{\omega} \cosh(\omega \tau) \;\;,\;\;
t(\tau)=\frac{z_0k}{\omega}
\sinh(\omega \tau)
\;\;,\;\;
x^{2}=x^{3}=\ldots=x^{D-2}=0.
\label{eqnxt}
\end{eqnarray}
Here $\omega = \sqrt{a^2-k^2}$, $z_0$ is a constant and $\tau$ is the proper time.
In the same spirit we can define another path in the $x^2-t$ direction and so on. However such paths in $x^1-t$ and $x^2-t$ planes are clearly related by spacetime isotropy. To be more precise all the $x_j$ directions\footnote{the index $j$ takes the value, $j=1,...., D-2$.} are related by rotational symmetry. But from the previous  equations \eqref{eq12} and \eqref{eqnxt} it is not very clear if the paths defined in $z-t$ and $x^1-t$ planes are also related by spacetime isotropy. If they are related by AdS spacetime isotropy then the detector response should be same along both the paths. In appendix $C$ we also explain how the paths in \eqref{eq12} and \eqref{eqnxt} can be related by spacetime isometry. But, whichever super critical path we choose either given by $x_j-t$ plane or $z-t$ plane the form of conformal dimension will always be given by \eqref{path2}. This actually dictate the form of the required $2n-$point correlator, upon which
response function is directly dependent.\\\\
Following eq. \eqref{two}, 
the two point function 
for uniform acceleration (in any supercrtical path) becomes,
\begin{eqnarray}
     G_{\rm AdS_{D}}(\Delta\tau) &=& \frac{\omega^{D-2}\Gamma(\frac{D}{2}-1)}{(4\pi)^{\frac{D}{2}}}\bigg(\frac{1}{i^{D-2}\sinh^{D-2}(\frac{\omega\Delta\tau}{2}-i\epsilon)}
    \nonumber \\
    & \ & \ \ \ -\frac{1}{(\sinh(A+(\frac{\omega\Delta\tau}{2}-i\epsilon)))^{\frac{D}{2}-1}(\sinh(A-(\frac{\omega\Delta\tau}{2}-i\epsilon)))^{\frac{D}{2}-1}}\bigg).\nonumber\\
\end{eqnarray}

Here $\sinh{A} = \omega/k$. Using the time translation invariance we can define the   transition
probability rate or detector's response function (per unit time)
for interaction Lagrangian \eqref{kukumia} of scalars\cite{Weinberg72},
\begin{eqnarray}
\mathcal{F}^{(n)}_{\rm AdS_{D}}=\int_{-\infty}^\infty d\Delta\tau e^{-iE\Delta \tau}
G^{(2n)}_{\rm AdS_{D}}
(\Delta\tau).\label{eqn:detrate}
\end{eqnarray}
Here, $G_{\rm AdS_{D}}^{(2n)}(\tau-\tau')=
\bra{0}:\Phi^n(x(\tau)):
:\Phi^n(x(\tau')):\ket{0}$ is the $2n$ correlator. For $n=1$ we obtain the response function defined in Jennings\cite{Jennings}, where the correlator becomes the two point correlator, i.e. the Wightman function.\\\\
It is well known that
KMS condition implies detailed
balance and thermality\cite{Jennings}. But just like  Minkowski spacetime, the fact that the KMS condition holds for Anti-de Sitter spacetime
does not specify the shape of the response spectrum for a particle detector but it can dictate us if the response function is proportional to Bose Einstein or Fermi Dirac distribution. We will describe the situation briefly.
Let us think of  a
worldline $x(\tau )$ generated by the timelike vector $\partial_t$
over which
a quantum field $\phi[x(\tau)]$ is considered.
If the correlator $\mathcal{G}$ of the field
obeys,
\begin{eqnarray}
  \mathcal{G}(\Delta \tau+i\beta)=(-1)^{2s}   \mathcal{G}(\Delta \tau)\label{22},\end{eqnarray}\\
then  $s=0$ and $s=\frac{1}{2}$ dictates periodic and anti-periodic respectively with $\beta $ being the inverse temperature. The response function is basically the Fourier of the correlator. Consequently if one takes
the Fourier transform
of the correlation function it becomes proportional to the Bose/Fermi distribution depending upon the periodic or anti-periodic condition respectively,
\begin{eqnarray}
    \hat{\mathcal{G}}\propto\frac{1}{e^{\beta E}-(-1)^{2s}}.
\end{eqnarray}
For more detailed discussion, interested readers are asked to look at \cite{Jennings} and page $6$ of ref. \cite{ottoEngine}.
Following such argument
we can now  claim the following-
\\\\\normalsize
\fbox{\begin{minipage}{45em}
\textbf{Theorem 1:}
 The detector response function
of massless scalar fields
coupled to the detector according to
eqn. \eqref{kukumia}
 for any uniform linear accelerated path for $(D > 2)$ dimensional AdS
spacetime defined   is equal to  \\ \\ - Bose-Einstein distribution (for even $D$ and any $n$). \\
- Bose-Einstein distribution (for odd $D$ and even $n$). \\
- Fermi-Dirac distribution (for odd $D$ and odd $n$).\\\\
multiplied by another  function dependent upon energy and temperature.
\end{minipage}}

Therefore we see that an inversion of statistics occurs in AdS spacetime for the third case. We now proceed to prove the statement.

\textbf{Proof:}
\noindent
In order to examine the proposed claim we investigate the KMS condition.
 The $2n$-point function $G^{(n)}\left({\tilde x},
{\tilde x'}\right)$ is related to the the Wightman function in the following way by Wick's theorem \cite{das},
\begin{equation}
G^{(n)}_{\rm AdS_{D}}\left({ x}, { x'}\right)
= \left(n!\right)\, \left(G_{\rm AdS_{D}}\left({ x}.
{ x'}\right)\right)^n.\label{eqn:2nptfnmv}
\end{equation}
So, the $2n$ correlator becomes,
\begin{multline}
G_{\rm AdS_{D}}^{(n)}(\Delta\tau)= (n!){\cal C}_{D}^{n} \bigg(
\frac{\omega}{\sqrt{2}k}\bigg)^{n(D-2)}\bigg(\frac{1}{i^{D-2}\sinh^{D-2}(\frac{\omega\Delta\tau}{2}-i\epsilon)}
    \\
    -\frac{1}{(\sinh(A+(\frac{\omega\Delta\tau}{2}-i\epsilon)))^{\frac{D}{2}-1}(\sinh(A-(\frac{\omega\Delta\tau}{2}-i\epsilon)))^{\frac{D}{2}-1}}\bigg)^n.
\label{eqn:2nptfnmvrind}
\end{multline}

This expression can be expanded using binomial series to the following form  
\begin{multline}
G_{\rm AdS_{D}}^{(n)}(\Delta\tau)= (n!){\cal C}_{D}^{n} \bigg(
\frac{\omega}{\sqrt{2}k}\bigg)^{n(D-2)} \sum^{n}_{\alpha = 0} \binom{n}{\alpha} \bigg(\frac{1}{i^{D-2}\sinh^{D-2}(\frac{\omega\Delta\tau}{2}-i\epsilon)}\bigg)^{n - \alpha}
    \\
    \bigg(\frac{-1}{(\sinh(A+(\frac{\omega\Delta\tau}{2}-i\epsilon)))^{\frac{D}{2}-1}(\sinh(A-(\frac{\omega\Delta\tau}{2}-i\epsilon)))^{\frac{D}{2}-1}}\bigg)^{\alpha}.
\label{eqn:binomial}
\end{multline}
 We rewrite the expression in the following form,
\begin{equation}
    G_{\rm AdS_{D}}^{(n)}(\Delta\tau)=(n!) {\cal C}_{D}^{n} \bigg(
\frac{\omega}{\sqrt{2}k}\bigg)^{n(D-2)} \sum^{n}_{\alpha = 0} \binom{n}{\alpha} \frac{(-1)^{\alpha}}{i^p} \
\ \mathcal{G}_{D,n,\alpha} (\rho)
\end{equation}
where
\begin{eqnarray}    &&\mathcal{G}_{D,n,\alpha} (\rho) = (\sinh(\rho-i\epsilon))^{-p} (\sinh(A + (\rho-i\epsilon)))^{-q} (\sinh(A-(\rho-i\epsilon)))^{-q}\\
&&\rho = \frac{\omega \Delta \tau}{2} \label{rhooo}\\
&&p = (n-\alpha)(D-2)\\
&& q = \alpha(\frac{D}{2} - 1).
\end{eqnarray}
Hence,\\
\begin{eqnarray}
    \mathcal{G}_{D,n,\alpha} (\rho + i \pi) & =& (\sinh(\rho + i \pi-i\epsilon))^{-p}(\sinh(A + (\rho + i \pi-i\epsilon)))^{-q}
    (\sinh(A-(\rho + i \pi-i\epsilon)))^{-q} \nonumber\\
 &=& \big( (-1) \sinh(\rho -i\epsilon)\big)^{-p} \big( (-1) \sinh(A + (\rho -i\epsilon))\big)^{-q} 
    \big( (-1) \sinh(A-(\rho -i\epsilon))\big)^{-q} \nonumber\\
&=&    (-1)^{-p-2q}(\sinh(\rho-i\epsilon))^{-p} (\sinh(A + (\rho-i\epsilon)))^{-q} (\sinh(A-(\rho-i\epsilon)))^{-q} \nonumber\\
&=&
   (-1)^{-n(D-2)}\mathcal{G}_{D,n,\alpha} (\rho ).
\end{eqnarray}

Now we can simply check the periodicity of   ${G}^{(n)}_{AdS_{D}}$,
\begin{eqnarray}
    {G}^{(n)}_{AdS_{D}}(\Delta\tau+\frac{2\pi\dot{i}}{\omega}) &=& (n!){\cal C}_{D}^{n} \bigg(\frac{\omega}{\sqrt{2}k}\bigg)^{n(D-2)} \sum^{n}_{\alpha = 0} \binom{n}{\alpha}  \frac{(-1)^{\alpha}}{i^p}  \mathcal{G}_{D,n,\alpha} (\rho+i\pi)\nonumber \\
    &=&(n!){\cal C}_{D}^{n} \bigg(\frac{\omega}{\sqrt{2}k}\bigg)^{n(D-2)} \sum^{n}_{\alpha = 0} \binom{n}{\alpha} \frac{(-1)^{\alpha}}{i^p}   (-1)^{nD}\mathcal{G}_{D,n,\alpha} (\rho) \nonumber\\
     & =& (-1)^{nD} {G}^{(n)}_{AdS_{D}}(\Delta\tau). \label{KMScondition}
\end{eqnarray}\\
In the case any of $D$ or $n$ (or both)
is even
the KMS condition is periodic and we will have we will have usual bosonic distribution.
According to Jennings\cite{Jennings}
for odd dimension the inversion of statistics happen but we have found a counterexample in the case $n$
is even.
So for odd $D$ but even $n$
we will have periodic KMS relation and as a result no inversion of statistics happen.
But we will always have antiperiodic condition  ${G}^{(n)}_{AdS_{D}}(\Delta\tau+\frac{2\pi\dot{i}}{\omega})=- {G}^{(n)}_{AdS_{D}}(\Delta\tau)$ for odd $n$ in odd dimension. 
\subsection{Detector response function for even dimensions}

\begin{figure}[t]
    \centering
    \includegraphics{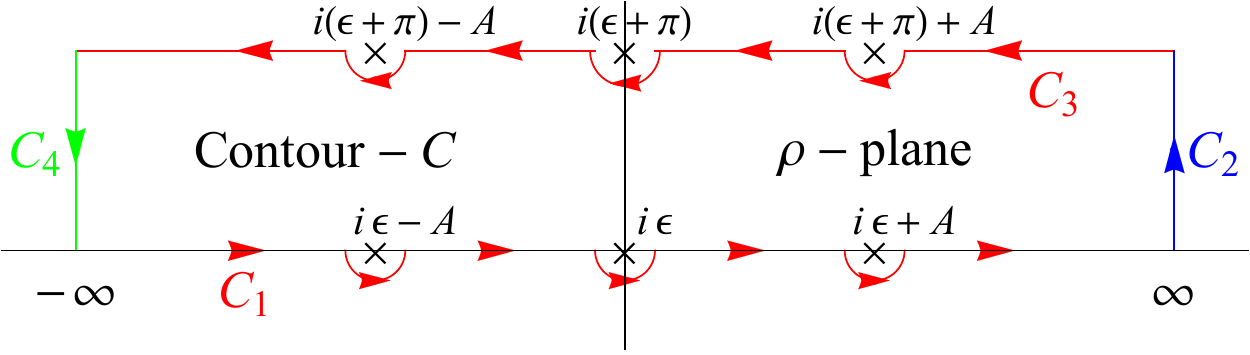}
    \caption{Contour $C$ for $I_{D,n,\alpha}$}.
    \label{fig:fig1}
\end{figure}

For any $D$ odd or even, $p$ is always an integer for any $n$ which is always integer as well. But when $D$ is odd, $q$ could be non integer which in turns create branch cut and this integral becomes notoriously difficult to do\cite{Jennings}.  In this section we calculate the detector response function for any $n$ in even dimension.
On substituting the $2n$-point
function~(\ref{eqn:2nptfnmvrind}) in the
expression~(\ref{eqn:detrate}) we get that,

\begin{equation}
\begin{split}
   \mathcal{F}_{AdS_D}^{(n)} & =\int_{-\infty}^\infty d\Delta\tau e^{-iE\Delta \tau} {\cal C}_{D}^{n} \bigg( \frac{\omega}{\sqrt{2}k}\bigg)^{n(D-2)} \sum^{n}_{\alpha = 0} \binom{n}{\alpha} \frac{(-1)^{\alpha}}{i^p} \ \ \mathcal{G}_{D,n,\alpha} (\rho) \\
    & = {\cal C}_{D}^{n} \bigg(\frac{\omega}{\sqrt{2}k}\bigg)^{n(D-2)} \sum^{n}_{\alpha = 0} \binom{n}{\alpha} \frac{(-1)^{\alpha}}{i^p} \ \int_{-\infty}^\infty \frac{2}{\omega} d\big( \frac{\omega \Delta\tau}{2} \big) e^{-i\big(\frac{2E}{\omega}\big) \frac{\omega \Delta \tau}{2}} \mathcal{G}_{D,n,\alpha} (\rho) \\
    & = {\cal C}_{D}^{n} \bigg(\frac{\omega}{\sqrt{2}k}\bigg)^{n(D-2)} \sum^{n}_{\alpha = 0} \binom{n}{\alpha} \frac{(-1)^{\alpha}}{i^p} \frac{2}{\omega} \ \int_{-\infty}^\infty d \rho
     e^{-i\Omega \rho} \ \mathcal{G}_{D,n,\alpha} (\rho)
\end{split}
\end{equation}
where $\Omega = \frac{2E}{\omega}$. Finally, denoting  the integral,

\begin{equation}
    F_{D,n,\alpha} = \int_{-\infty}^\infty d \rho \
     e^{-i\Omega \rho} \ \mathcal{G}_{D,n,\alpha} (\rho)\label{response}.
\end{equation}

This integral
has poles of order $p$ and $q$ all along the imaginary axis at the points $\rho = i(\epsilon - 2m\pi ) $ and $\rho = i(\epsilon - 2m\pi ) \pm A$ respectively with $m\in \mathbb{Z}$.

Now for integer values of $p$ and $q$, we define a closed contour integral $I_{D,n,\alpha}$ where the contour $C$ is given in figure \ref{fig:fig1}. This contour contain a pole located at $\rho = i \epsilon \approx 0$ and $\rho = i \epsilon \pm A.$

\begin{equation}
\begin{split}
    I_{D,n,\alpha} & = \oint_C d \rho \
     e^{-i\Omega \rho} \ \mathcal{G}_{D,n,\alpha} (\rho) \\
    & = \int_{C_1} +\int_{C_2} + \int_{C_3} + \int_{C_4} d \rho \
     e^{-i\Omega \rho} \ \mathcal{G}_{D,n,\alpha} (\rho).
\end{split}\label{alcontur}
\end{equation}

 For the contour $C_2$ the real part of $\rho$ is $\infty$. According to equation (\ref{rhooo}) if $\rho \to \infty$, then $\Delta \tau \to \infty$. Thus from equation (\ref{eqn:2nptfnmvrind}) these imply that $\mathcal{G}_{D,n,\alpha} (\rho) \to 0$.  Therefore the integral for contour $C_2$ vanishes because denominator goes to infinity.
 \begin{eqnarray}
    \int_{C_2} d \rho \
     e^{-i\Omega \rho} \ \mathcal{G}_{D,n,\alpha} (\rho)  & = & \int_{\rho = \infty + i\epsilon}^{\rho = \infty + i(\epsilon + \pi)} d \rho \
     e^{-i\Omega \rho} \mathcal{G}_{D,n,\alpha} (\rho) \nonumber \\ 
     &=& \int_{ \sigma = \rho - \infty = i\epsilon}^{\sigma = \rho - \infty = i(\epsilon + \pi)} d \rho \
     e^{-i\Omega \sigma} e^{-i\Omega \infty} \mathcal{G}_{D,n,\alpha} (\sigma + \infty) = 0.
\end{eqnarray}

This is also true for the integral with contour $C_4$. However, for contour $C_3$ we get

\begin{equation}
\begin{split}
    \int_{C_3} d \rho \
     e^{-i\Omega \rho} \ \mathcal{G}_{D,n,\alpha} (\rho)  & = \int_{\rho = \infty + i(\epsilon + \pi)}^{\rho = -\infty + i(\epsilon + \pi)} d \rho \
     e^{-i\Omega \rho} \ \mathcal{G}_{D,n,\alpha} (\rho)\\
     & = \int_{\sigma = \rho - i\pi = \infty + i\epsilon }^{\sigma = \rho - i\pi = -\infty + i\epsilon} d \sigma \
     e^{-i\Omega \sigma} e^{\Omega \pi} \ \mathcal{G}_{D,n,\alpha} (\sigma + i \pi).
\end{split} \label{c3eqn}
\end{equation}


Putting this into eq. (\ref{c3eqn}), we obtain that,
\begin{equation}
    \begin{split}
        \int_{C_3} d \rho \
     e^{-i\Omega \rho} \ \mathcal{G}_{D,n,\alpha} (\rho)  & = \int_{\sigma= \infty + i\epsilon }^{\sigma =-\infty + i\epsilon} d \sigma  e^{-i\Omega \sigma} e^{\Omega \pi} (-1)^{n(D-2)}\mathcal{G}_{D,n,\alpha}(\sigma) \\
     & =  \frac{-e^{\Omega \pi}}{(-1)^{n(D-2)}} \int_{\sigma= -\infty + i\epsilon }^{\sigma =\infty + i\epsilon} d \sigma  e^{-i\Omega \sigma}  \mathcal{G}_{D,n,\alpha}(\sigma) \\
     & =  \frac{-e^{\Omega \pi}}{(-1)^{n(D-2)}} F_{D,n,\alpha}.
    \end{split}
\end{equation}

Finally, we get the relationship between eq. (\ref{response}) and (\ref{alcontur}),
\begin{equation}
    F_{D,n,\alpha} = \frac{I_{D,n,\alpha}}{1-\frac{e^{\Omega \pi}}{(-1)^{n(D-2)}}} = \frac{-(-1)^{n(D-2)} I_{D,n,\alpha}}{e^{\Omega \pi} - (-1)^{n(D-2)}}. \label{statinv}
\end{equation}
\begin{figure}[t]
    \centering
    \includegraphics[scale=0.7]{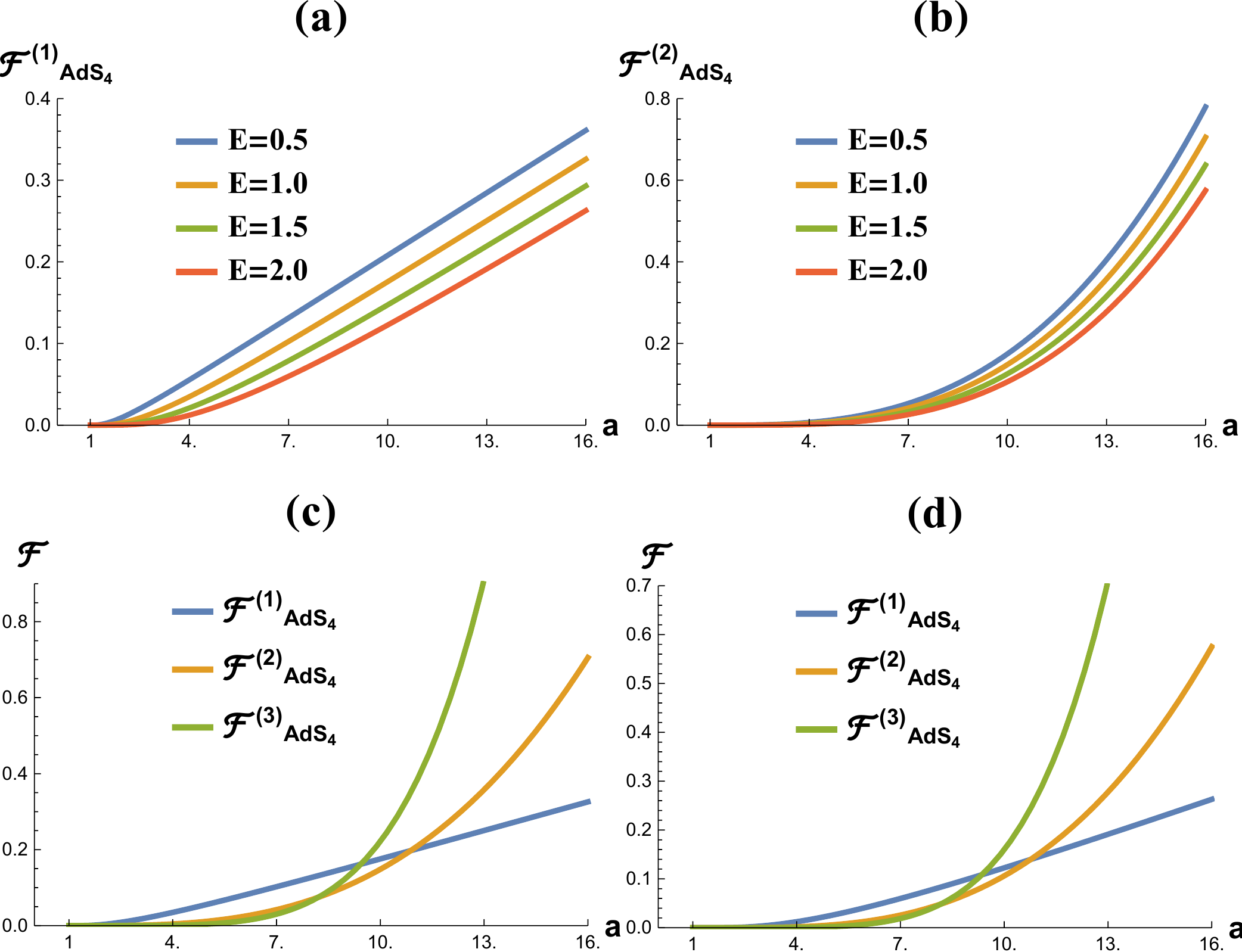}
    \caption{Behaviour of  detector response functions in (a) $\mathcal{F}^{(1)}_{AdS_4}$ and (b) $\mathcal{F}^{(2)}_{AdS_4}$ as a function of acceleration $a$ for different values of energy $E$.  Comparison  of $\mathcal{F}^{(1)}_{AdS_4}$, $\mathcal{F}^{(2)}_{AdS_4}$ and $\mathcal{F}^{(3)}_{AdS_4}$   for (c) $E=4$  and (d) $E=8$ as we change the  acceleration. In this plot we fixed AdS radius $L=1$.}.
    \label{fig:fig3}
\end{figure}
\begin{figure}[t]
    \centering
    \includegraphics[scale=0.7]{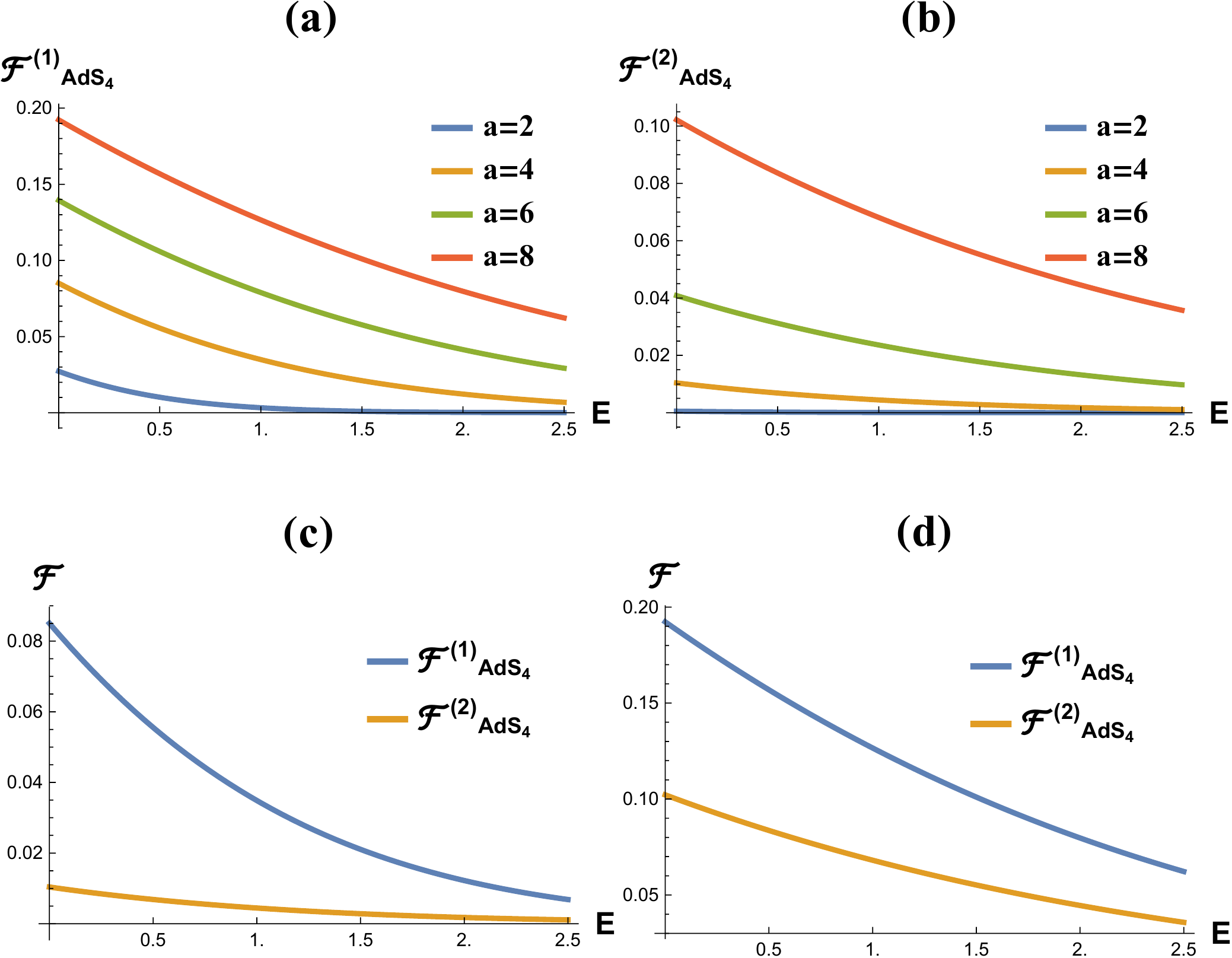}
    \caption{Behaviour of  detector response functions in (a) $\mathcal{F}^{(1)}_{AdS_4}$ and (b) $\mathcal{F}^{(2)}_{AdS_4}$ as a function of energy $E$ for different values of acceleration $a$.  Comparison  of $\mathcal{F}^{(1)}_{AdS_4}$ and
    $\mathcal{F}^{(2)}_{AdS_4}$
    for (c) $a=4$  and (d) $a=8$ as we change the  energy. In this plot we have also set AdS radius to unity.}.
    \label{fig:fig2}
\end{figure}
So the expression for $I_{D,n,\alpha}$ can be found using contour integration,
\begin{equation}
\begin{split}
    I_{D,n,\alpha} & =  2\pi i \times \{ \text{sum of the residues at } \rho=0 , \pm A \text{ of } \mathcal{G}_{D,n,\alpha}(\rho) \} \\
    & =  2\pi i \times \bigg( \lim_{\rho \to 0} \frac{\Theta(p-1)}{\Gamma(p)} \bigg( \frac{1}{\cosh \rho} \frac{d}{d\rho} \bigg)^{p-1} \frac{e^{-i\Omega\rho}}{\sinh^q{(A+\rho)} \cosh {(\rho)} \sinh^q{(A-\rho)}} + \\
    & \ \ \ \ \lim_{\rho \to -A} \frac{\Theta(q-1)}{\Gamma(q)} \bigg( \frac{1}{\cosh (\rho+A)} \frac{d}{d\rho} \bigg)^{q-1} \frac{e^{-i\Omega\rho}}{\cosh{(A+\rho)} \sinh^p{(\rho)} \sinh^q{(A-\rho)}} +  \\
    & \ \ \ \ \lim_{\rho \to A} \frac{\Theta(q-1)}{\Gamma(q)} \bigg( \frac{-1}{\cosh (A-\rho)} \frac{d}{d\rho} \bigg)^{q-1} \frac{-e^{-i\Omega\rho}}{\sinh^q{(A+\rho)} \sinh^p {(\rho)} \cosh{(A-\rho)}} \bigg).
\end{split}.
\end{equation}
Here, $\Theta(p)$ is Heaviside step function and $\Gamma(p)$ is gamma function. Consequently, we have
\begin{equation}
   \mathcal{F}_{AdS_D}^{(n)} = \Bigg(n! \ {\cal C}_{D}^{n} \bigg(\frac{\omega}{\sqrt{2}k}\bigg)^{n(D-2)}
    \sum^{n}_{\alpha = 0} \binom{n}{\alpha} \frac{(-1)^{n(D-2)+\alpha+1}}{i^p} \frac{2}{\omega} \  I_{D,n,\alpha}\Bigg)\frac{1}{e^{2\pi E /\omega} - (-1)^{n(D-2)}}. \label{responseFunction}
\end{equation}
\begin{figure}[t]
    \centering
    \includegraphics[scale=0.7]{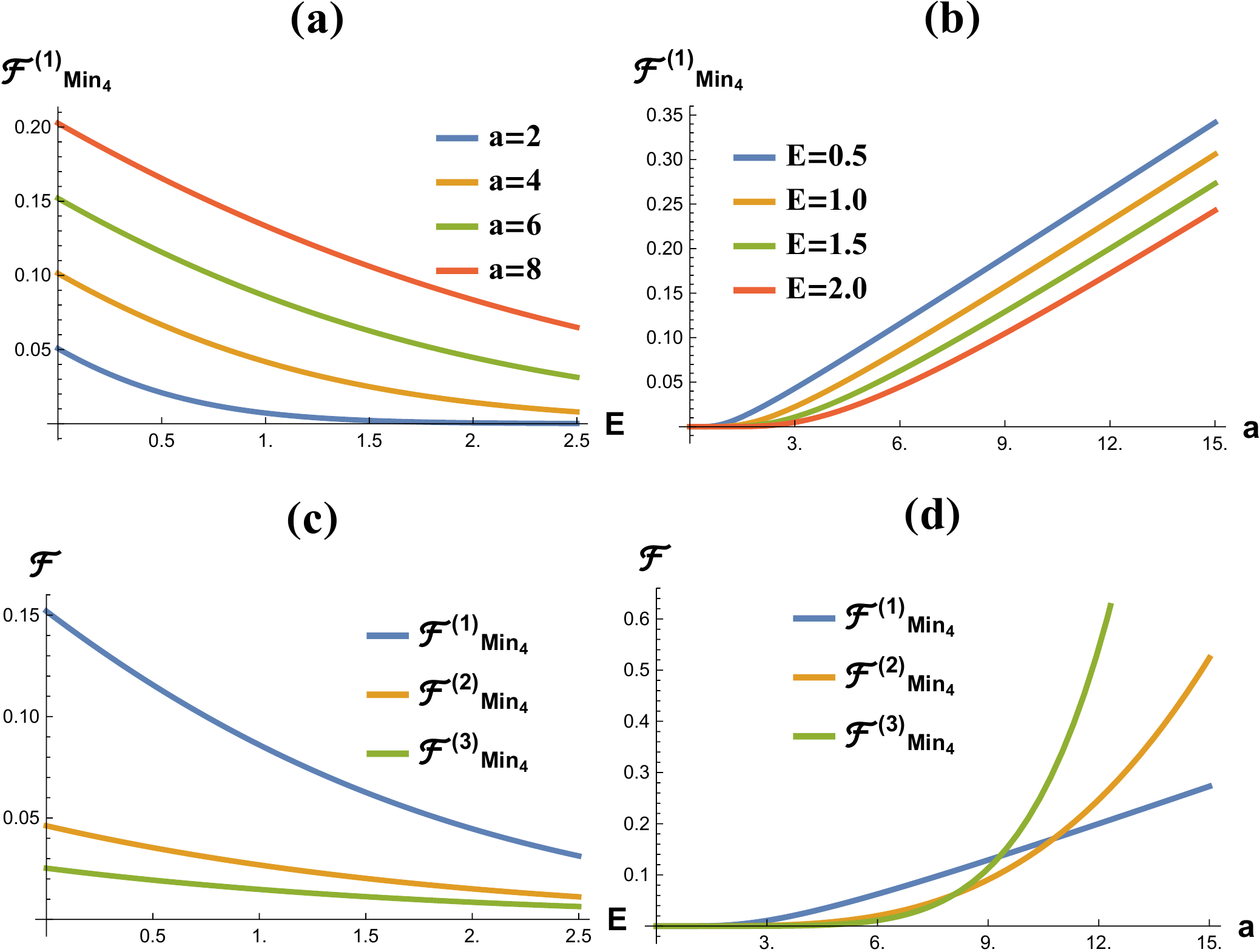}
    \caption{
    Characteristics of four dimensional detector response function in flat spacetime with $n=1$ as (a) a function of energy with different acceleration  and (b) as a function of acceleration  with different energy. Comparison on the effects of non-linearity on $\mathcal{F}_{Min_4}$
    as function of (c) energy and (d)  acceleration.
    }
    \label{fig:fig4}
\end{figure}
Thence we can find out that the KMS condition is maintained at a temperature\cite{Jennings},
\begin{eqnarray}
    T=\frac{\sqrt{a^2-k^2}}{2\pi}.
\end{eqnarray}
Using \eqref{responseFunction}, we can calculate explicit expression for different values of $D$ and $n$.
For example, we take $n=1$ and $D=4$ we have \cite{Jennings},
\begin{equation}
    \mathcal{F}_{AdS_4}^{(1)} = \Bigg(\frac{E}{2\pi}-\frac{k^{2}}{4\pi a}\sin\bigg(\frac{2E}{\omega}\sinh^{-1}(\frac{\omega}{k})\bigg)\Bigg)\frac{\Theta(a-k)}{e^{2\pi E/\omega}-1}.
\end{equation}
Also looking into $n=2,$ $3$ in four dimensions,
\begin{multline}
    \mathcal{F}_{AdS_4}^{(2)} = \bigg( \ \frac{\text{E} \left(a^2+\text{E}^2-4 k^2\right)}{24 \pi ^3} 
    -\frac{k^4 \left(k^2-6 a^2\right)}{64 \pi ^3 a^3} \sin \left( \frac{2 E}{\omega} \sinh^{-1}\left(\frac{\omega}{k }\right)\right)- \\ 
    \frac{\text{E} k^4}{32 \pi ^3 a^2} \cos \left( \frac{2 E}{\omega} \sinh^{-1}\left(\frac{\omega}{k }\right)\right) \bigg)
    \frac{\Theta(a-k)}{e^{2\pi E/\omega}-1}.
\end{multline}
\begin{multline}
    \mathcal{F}_{AdS_4}^{(3)}=\frac{1}{20480 \pi ^5 a^5}\bigg( 15 k^6 \big( 6 a E \left(6 a^2-k^2\right) \cos\left(\frac{2 E }{\omega }\sinh ^{-1}\left(\frac{\omega }{k}\right)\right)+\left(-80 a^4+4 a^2 \left(E^2+5 k^2\right)-3 k^4\right) \\ \sin \left(\frac{2 E }{\omega }\sinh ^{-1}\left(\frac{\omega }{k}\right)\right)\big)+ 
    32 a^5 E \left(4 a^4+a^2 \left(5 E^2-23 k^2\right)+E^4-20 E^2 k^2+64 k^4\right) \bigg)   \frac{\Theta(a-k)}{e^{2\pi E/\omega}-1}.
\end{multline}\\\\\\
We can also 
obtain explicit expression
for any $n$ for other dimensions that $D=4$ but we concentrate on $D=4$ at this moment. 
Let us examine the effect of non linearity on
$\mathcal{F}^{(n)}_{AdS_4}$
in four dimensions 
with $n=1,$ $2$ and $3$. We first show  the relation between detector response function   and  acceleration. In figure 2 (a) and 2(b) we plot $\mathcal{F}^{(1)}_{AdS_4}$
and $\mathcal{F}^{(2)}_{AdS_4}$ separately  with respect to $a$ for different energy $E$. As expected we find out when
acceleration is increased and as a result 
temperature  is enhanced more particles are detected in the detector.
Same trend is also noted for any $\mathcal{F}^{(n)}_{AdS_4}$. 
It should be noted  that
there is a 
a difference in between
 the
particle content deduced through the "clicking" of any detector
and
the usual notion of Fock-space  of particle content\cite{Ted Jacobson,last,Jennings}. 
In the most general case these approaches differ from each other
for the most general trajectory
but coincides for a constantly accelerated observer\cite{Ted Jacobson,last,Jennings}.
Next 
in $2$ $(c)$ and $2$ $(d)$,
we have compared how the  $\mathcal{F}^{(n)}_{AdS_4}$ with $n=1,$ $2$ and $3$ changes with acceleration and energy. We can see both at energy $E= 4$ and $E=8$ detector response grow linearly with temperature with $n=1$. On the other hand with $n=2$ and $n=3$ we see  $\mathcal{F}_{AdS_4}$ changes in a non-linear fashion as we increase the acceleration.
We  have also seen the same  pattern for other values of $n$ that, as we increase $n$ the detector response changes more drastically with temperature. Next we analysed
the response function with energy in figure (3). 
We can see that for any
constant 
acceleration $a$ the detector detects fewer particles with higher energy. Detector response function in AdS spacetime also depends upon the AdS radius $L=\frac{1}{k}$ with $k$ being the curvature
of the spacetime. 
Putting  curvature $k \to 0$ in the correlation function (\ref{responseFunction}).
Taking the limit $k\rightarrow0$
of our Wightman function for AdS
one can obtain the two Wightman function in Minkowski spacetime
(see eq. $(21)$ of ref.\cite{Jennings}).
As a result 
we can recover the response function for $D$-dimensional Minkowski spacetime. We can carry out the calculation by simply calculating only $\alpha = 0$ term in equation (\ref{responseFunction}). 
Thus, for both even or odd $D$ dimensions we obtain the response function in Minkowski spacetime,
\begin{multline}
    \mathcal{F}_{Min_{D}}^{(n)} =  \left( \frac{(-a)^{D-2}\Gamma(D/2-1)}{(4 \pi)^{D/2}  \ i^{D-2}}\right)^{n} \times \\ \ 2 \pi i \times \bigg( \lim_{\rho \to 0} \frac{-1}{\Gamma(n(D-2))} \bigg( \frac{1}{\cosh \rho} \frac{d}{d\rho} \bigg)^{n(D-2)-1} \bigg(\frac{e^{-i\Omega\rho}}{\cosh {(\rho)} }\bigg) \bigg)\frac{n!}{e^{2\pi E/a}-(-1)^{nD}}.\label{lop}
\end{multline}\\

Using \eqref{lop}  we can also acquire
explicit expression $D=4$ and $n=1,2$ and $3$ as well,
\begin{eqnarray}
    \mathcal{F}_{Min_{4}}^{(1)} &=& \frac{E}{2 \pi }\frac{1}{e^{2\pi E/a}-1}, \label{flat1} \\
    \mathcal{F}_{Min_{4}}^{(2)} &=& \frac{E \left(a^2+E^2 \right)}{24 \pi ^3}\frac{1}{e^{2\pi E/a}-1},\\
    \mathcal{F}_{Min_{4}}^{(3)} &=& \frac{E \left(4 a^2+E^2 \right) \left(a^2+E^2 \right)}{640 \pi ^5}\frac{1}{e^{2\pi E/a}-1}, \label{flat3}
\end{eqnarray}\\
 In ref.\cite{ottoEngine} detector response function for quadratic and linear scalar coupling is well studied but we have a general formula from which one can study detector response for arbitrary non-linearity in Minkowski spacetime. Interested readers can follow the link at \cite{dropbox} in order access the mathematica file. One can just obtain the exact expressions of the  detector response using it for any $n $  using eq. (\ref{responseFunction}) and (\ref{lop}) in AdS and Minkowski spacetime as well.
 Now,
the results of theorem 1 also holds in flat spacetime\cite{sri} but using the results of \eqref{flat1} to \eqref{flat3}, we can now compare the affects of non linearity for flat spacetime.
Following the previous example of AdS spacetime we also have plotted the dependence of response function on  temperature and energy.
\section{Dirac Fields in AdS}
Turning our attention towards  Dirac fermions in AdS spacetime minimally coupled to 
background gravity,
the action is-
\begin{eqnarray}
    \mathcal{S}_{0}=\int d^Dx\sqrt{|g|}
    \bar\Psi
    i\slashed{D} \Psi.\label{ld}
\end{eqnarray}

   Before Here we choose a local Lorentz frame (vielbein) which is defined as,
\begin{eqnarray}
    e^a_\mu=\frac{\delta^a_\mu}{kz},
\end{eqnarray} such that \begin{eqnarray}
    g_{\mu\nu}=e^a_\mu e^b_\nu \eta_{ab},
\end{eqnarray} depicts the spacetime metric before. Here Latin letters
\(a, b\) corresponds to local orthonormal coordinates and Greek letters
\(\mu, \nu\) signifies the spacetime coordinates, both of them run from
\(0\) to \(D-1\). Also $\eta_{ab} = \text{diag}(+1,-1, ~\ldots{} ~,
-1) $ is the local flat metric. Moreover, \begin{eqnarray}
     e^\mu_a e^b_\mu = \delta^b_a, \nonumber\\
     e^\mu_a e^a_\nu = \delta^\mu_\nu.
 \end{eqnarray} Now, the curved space \(\Gamma\) matrices and the
covariant derivatives are defined as, \begin{eqnarray}
   && \Gamma^\mu= e_{a}^\mu\gamma^a,\nonumber\\
&& D_\mu=    \partial_{\mu} + \frac{1}{2}\omega^{bc}_{\mu} \Omega_{bc} 
\end{eqnarray} \(\gamma_a\) are flat spacetime gamma matrices \footnote{For similcity in notation we are going to use $z$ for $x^{D-1}$ and the index $\mu = z$ instead of $\mu = D-1$.}. Here
\(\Omega_{bc}\) is commutator between \(\gamma\) matrices,
\begin{eqnarray}
    \Omega^{bc} = \frac{1}{4}(\gamma^b \gamma^c - \gamma^c \gamma^b)
\end{eqnarray} and the spin connections \(\omega^{bc}_{\mu}\) are,
\begin{eqnarray}
    \omega^{ab}_{\mu}=e^{a \lambda }\big(\partial_\mu e^{ b}_{\lambda}-\Gamma^\alpha_{\mu\lambda}e^{b}_{\alpha}\big)\label{spinConn}.
\end{eqnarray} and, \(\Gamma^\nu_{\sigma\mu}\) are the Christoffel
symbols related to AdS spacetime metric eq. (1). Here \(\Gamma^{\mu}\)
and \(\gamma^a\) satisty the following Clifford algebra,
\begin{eqnarray}
    \{ \Gamma^{\mu}, \Gamma^{\nu} \} &=  2g^{\mu \nu} \mathbb{I}_{N \times N}  \nonumber\\
     \{\gamma^{b}, \gamma^{c} \} &= 2 \eta^{b c} \mathbb{I}_{N \times N}.  \label{diracMin}
\end{eqnarray} with, \begin{eqnarray}
N =
  \begin{cases}
    2^{\frac{D}{2}} & \text{$D$ is even} \\
    2^\frac{D-1}{2} & \text{$D$ is odd}. 
  \label{chandumama}  \end{cases}    
\end{eqnarray} The Dirac operator takes the following form in AdS,
\begin{eqnarray}
    \Gamma^{\mu}{D_{\mu}}  \equiv e^{\mu}_{a}\gamma^{a}\big( \partial_{\mu} + \frac{1}{2}\omega^{bc}_{\mu} \Omega_{bc} \big)  = k\bigg( \gamma^a \partial_{a} - \frac{D-1}{2} \gamma^z \bigg) .
\end{eqnarray}
We are considering 
 the DeWitt
detector coupled to
the Dirac field via the Lagrangian
 \cite{Nottingham,ottoEngine},
 \begin{eqnarray}
    \mathcal{L}_{int} = c  m(\tau)  \bar\Psi(x(\tau)) \Psi(x(\tau))   \label{interHam2}
 \end{eqnarray}
 where $c$ is coupling constant and $m(\tau)$ is the detector monopole moment operator as previous section.
 We can decompose the $\Psi$ field in positive and negative frequency part.
 \begin{eqnarray}
     \Psi(x)=\Psi^{+}(x)+\Psi^{-}(x).
 \end{eqnarray}
 The Wightman functions of the fermionic feld are,
 \begin{eqnarray}
     \left[S^{+}(x,x')\right]_{ab}=  \bra{0} \Psi_a(x)\overline{\Psi}_b(x')\ket{0}\\
      \left[S^{-}(x,x')\right]_{ab}=  \bra{0} \overline{\Psi}_b(x')\Psi_a(x)\ket{0}.
 \end{eqnarray}
Here $a$ and $b$ are spinor indices. We know the detector response function of
  fermions (per unit time)
 for interaction Lagrangian is  given by\cite{Weinberg72}, 
 \begin{equation}
    \mathcal{J}_{AdS_D}=\int_{-\infty}^\infty d\Delta\tau e^{-iE\Delta \tau} S^{(2)}_{D} (\Delta\tau) \label{chhapu}
 \end{equation}
 where,
 \begin{eqnarray}
     S^{(2)}_{D} (x(\tau),x(\tau')) &=& \bra{0}:\big( \overline{\Psi}_a(x(\tau))) \Psi_a(x(\tau)) \big)::\big( \overline{\Psi}_b(x(\tau'))) \Psi_b(x(\tau'))  \big):\ket{0}\nonumber\\ &=& \text{Tr}[S^{+}(x,x')S^{-}(x',x)]. \label{traces}
 \end{eqnarray}
 is 4-points correlator of Fermionic field.
 See eq. (B12) in reference \cite{ottoEngine} for more details. Demanding the following regularity and boundary condition\cite{adscft,spinorExchange}, 
 \begin{eqnarray}
 \lim_{z\to \infty} S(x,x')
     =
      \lim_{z'\to \infty} S(x,x')
      =0,\\
     \lim_{z\to 0} S(x,x')
     =
      \lim_{z'\to 0} S(x,x')
      =0.
 \end{eqnarray}\\ 
 The two point function for
 massless fermions in AdS spacetime takes the following form (see appendix for the detailed calculation)
 \cite{
 adscft,
 spinorExchange},
\begin{eqnarray}
S^+(x,x')  = i\sqrt{\frac{z'}{z}} \left(\Gamma^{\mu}D_{\mu} + \frac{\Gamma^z}{2z}  \right)  
     \left( \mathcal{P}^{+}G_{AdS_D}(x,x',+\frac{1}{2}) + \mathcal{P}^{-}G_{AdS_D}(x,x',-\frac{1}{2})
     \right)
\end{eqnarray} where, $P^{\pm}= (1 \pm i\gamma^z)/2$ and   $G_{AdS_D}$ is as usual,
\begin{eqnarray}
    G_{AdS_D}(x,x',\pm \frac{1}{2}) = \frac{k^{D-2}\Gamma\left(\frac{D-2}{2}\right)}{2(2\pi)^{D/2}} \left((v-1)^{\frac{2-D}{2}} \mp (v+1)^{\frac{D-2}{2}} \right). \label{scalarGGG}\nonumber
\end{eqnarray}\\
Following \eqref{sminus} the $S^-$ is related to, $S^{+}=-S^-$.\\\\
\normalsize	
\fbox{\begin{minipage}{45em}
\textbf{Theorem 2:}\\
    
    The response function of an uniform linearly accelerated 
    Unruh-DeWitt
    detector  quadratically coupled to a massless Dirac field in AdS vacuum in $D \geq 2$ spacetime dimensions equals
     the response function of a  detector coupled linearly and conformally to a massless scalar field in 
     $2D$ dimensional
     AdS vacuum   times     a numeric factor of   $ \frac{N \ (\Gamma(D/2))^2}{\Gamma(D-1)}$.
   
    \end{minipage}}\\
    
    This theorem was first described for flat space-time by Louko and Toussaint \cite{Nottingham}. We have generalised the result in 
    the context of
    for fermions
in    
    AdS spacetime.
    We can now proceed to prove the above statement.\\
\\
\textbf{Proof:}
  Re-writing the fermion propagator as, 
\begin{eqnarray}
    S^{+}(x,x') 
    &=& i\sqrt{\frac{z'}{z}} \bigg( \ (\Gamma^{\mu}  \partial_\mu u) {\xi}' (x,x') -\frac{D-2}{2z}\Gamma^z {\xi} (x,x')\bigg).  \label{Splus}
\end{eqnarray} where,
\begin{eqnarray}
    {\xi} (x,x') &=& \left( \mathcal{P}^{+}G_{AdS_D}(x,x',+\frac{1}{2}) + \mathcal{P}^{-}G_{AdS_D}(x,x',-\frac{1}{2})
     \right) \\
    &=& \mathcal{C}_D\left({(v-1)^{\frac{2-D}{2}}+i\gamma^z (v+1)^{\frac{2-D}{2}}}\right).
\end{eqnarray}
Therefore \({\xi}'(x,x')\) 
takes the following form,
\begin{eqnarray}
    {\xi}'(x,x')= \frac{\partial {\xi}(x,x')}{\partial u}=-\mathcal{C}_D\left(\frac{D-2}{2}\right)\left({u^{-\frac{D}{2}}+i\gamma^z (u+2)^{-\frac{D}{2}}}\right)
\end{eqnarray} with \(u=v-1\). As a result, \begin{eqnarray}
u =  -\frac{\eta_{a b}(x^a - x'^a)(x^b - x'^b)}{2zz'}.
\end{eqnarray}
Also, \begin{eqnarray}
    \Gamma^{\mu}  \partial_\mu u &=& \gamma^{a} e^{\mu}_{a} \partial_\mu \big( -\frac{\eta_{a b}(x^a - x'^a)(x^b - x'^b)}{2zz'} \ \big) \nonumber\\
    &=& \gamma^{a} (kz)  \big( \frac{\eta_{ab}(x'^b - x^b)}{zz'} -\frac{u\delta^z_\mu}{z} \big) \nonumber\\ 
    &=& k \big( \frac{\gamma^{a} \eta_{a b}(x'^b - x^b)}{z'} -{u\gamma^{z}} \big) \label{eqn89}.
\end{eqnarray} Next we put eqn. (\ref{eqn89}) to eqn. (\ref{Splus}) to 
obtain, \begin{eqnarray}
    S^{+}(x,x') = -ik\sqrt{\frac{z'}{z}} \bigg( \ \frac{\gamma^{c} \eta_{c b}(x^b - x'^b)}{z'} {\xi}'(x,x') + \gamma^z \big( u {\xi}'(x,x') + \frac{D-2}{2} {\xi}(x,x')\big)\bigg). \label{eqn90}
\end{eqnarray} 

Using eqn. (\ref{eqn90}) we can write,
\begin{eqnarray}
    S^{(2)}_{D}(x,x') &=& \text{Tr}[S^{+}(x,x')S^{-}(x',x)] \nonumber \\
    &=&  \frac{k^2}{zz'}{ \eta_{c b}(x^b - x'^b)} { \eta_{d a}(x'^a - x^a)} \text{Tr}[\gamma^{c}\xi'\gamma^{d}\xi'] \nonumber\\
    & & +\frac{k^2}{z'} { \eta_{c b}(x^b - x'^b)}\left(u \  \text{Tr}[\gamma^{c}\xi'\gamma^{z}\xi'] + \left(\frac{D-2}{2}\right) \text{Tr}[\gamma^{c}\xi'\gamma^{z}\xi] \right) \nonumber\\
    & & +\frac{k^2}{z} { \eta_{d a}(x'^a- x^a)}\left(u \  \text{Tr}[\gamma^{z}\xi'\gamma^{d}\xi'] + \left(\frac{D-2}{2}\right) \text{Tr}[\gamma^{z}\xi'\gamma^{d}\xi] \right) \nonumber\\
    & & +k^2\left( u^2 \ \text{Tr}[\gamma^{z}\xi'\gamma^{z}\xi'] + 2u \left(\frac{D-2}{2}\right) \text{Tr}[\gamma^{z}\xi'\gamma^{z}\xi]+\left(\frac{D-2}{2}\right)^2\text{Tr}[\gamma^{z}\xi\gamma^{d}\xi]
    \right). \label{4point}
\end{eqnarray} \\
We need to evaluate the traces. Let us first write the individual trace formulas.\\
\begin{eqnarray}
    \text{Tr}[\gamma^{c}\xi'\gamma^{d}\xi'] &=& \left(\frac{D-2}{2}\right)^2 \mathcal{C}^2_D \left( \frac{\text{Tr}[\gamma^c \gamma^{d}]}{u^{D}} + i \frac{\text{Tr}[\gamma^c \gamma^z\gamma^{d}]+\text{Tr}[\gamma^c \gamma^{d} \gamma^z]}{(u^2+2u)^{D/2}} - \frac{\text{Tr}[\gamma^c \gamma^z\gamma^{d}\gamma^z]}{(u+2)^{D}} \right) \nonumber \\
    &=& \left(\frac{D-2}{2}\right)^2 \ N \  \mathcal{C}^2_D \left( \frac{\eta^{c d}}{u^{D}} - \frac{2\eta^{c z} \eta^{z d}+ \eta^{c d}}{(u+2)^{D}} \right),\\
    \text{Tr}[\gamma^{c}\xi'\gamma^{z}\xi'] &=& \text{Tr}[\gamma^{z}\xi'\gamma^{c}\xi'] = \eta^{c z} \left(\frac{D-2}{2}\right)^2 \ N \  \mathcal{C}^2_D (u^{-D}+(u+2)^{-D}), \\
    \text{Tr}[\gamma^{c}\xi' \gamma^{z}\xi], &=& \text{Tr}[\gamma^{z}\xi'\gamma^{c}\xi] = -\eta^{c z} \left(\frac{D-2}{2}\right) \ N \  \mathcal{C}^2_D (u^{1-D}+(u+2)^{1-D}), \\
    \text{Tr}[\gamma^{z}\xi \gamma^{z}\xi]&=& - \ N \  \mathcal{C}^2_D (u^{2-D}+(u+2)^{2-D}).
\end{eqnarray}


In deriving the above equations we have
used the following identities- \begin{eqnarray}
    && \gamma^z \xi = \xi \gamma^z, \gamma^z \xi' = \xi' \gamma^z, \\
    &&\gamma^c \gamma^d + \gamma^d \gamma^c = 2 \eta^{c d} \mathbb{I}_{N \times N},\nonumber\\
&&Tr[\gamma^d \gamma^c] = N \eta^{c d}.\nonumber
\end{eqnarray} 
Using those traces equation (\ref{4point}) becomes
\begin{eqnarray}
    S^{(2)}_{D}(x,x') &=& k^2\ N \ \mathcal{C}^2_{D} \left(\frac{D-2}{2}\right)^2 \Bigg(  \frac{2u}{u^D}-\frac{2u}{(u+2)^D} + \frac{(z-z')^2}{zz'}\frac{2}{(u+2)^D}\nonumber \\
    && +\left(\frac{z-z'}{z'}+\frac{z'-z}{z'} \right)\left( \frac{u}{u^{D}}+\frac{u}{(u+2)^{D}} - \frac{1}{u^{D-1}}-\frac{1}{(u+2)^{D-1}}    \right) \nonumber \\
    && - \frac{u^2}{u^{D}}-\frac{u^2}{(u+2)^{D}} + \frac{2u}{u^{D-1}}+\frac{2u}{(u+2)^{D-1}} - \frac{1}{u^{D-2}}-\frac{1}{(u+2)^{D-2}} \Bigg) \nonumber \\
    &=& k^2\ N \ \mathcal{C}^2_{D} \left(\frac{D-2}{2}\right)^2 \left( \frac{2}{u^{D-1}} - \frac{2}{(u+2)^{D-1}} \right) \nonumber \\
     &=& N \frac{(\Gamma(D/2))^2}{\Gamma(D-1)}G_{AdS_{2D}}(x,x')
\end{eqnarray}
\begin{figure}[t]
    \centering
    \includegraphics[scale=0.7]{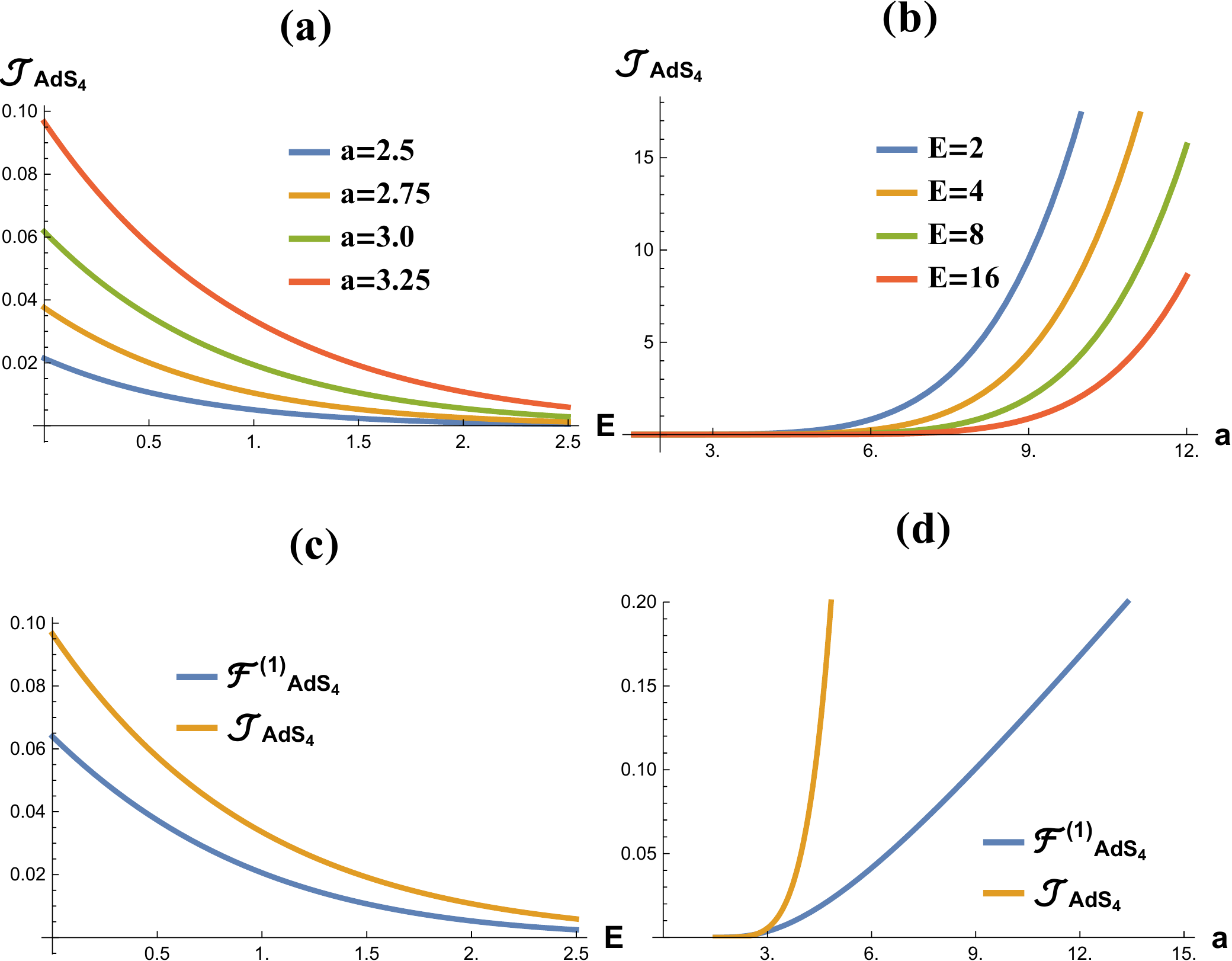}
    \caption{
    Plot (a) denotes fermionic detector response function $\mathcal{J}_{AdS_4}$ as a function of energy with different acceleration  and (b) as a function of acceleration  with different energy. Finally, we compare the scalar $\mathcal{F}^{(1)}_{AdS_4}$ and fermionic $\mathcal{J}_{AdS_4}$ response functions for specific value of $a = 3.25$ (plot c) and $E = 2$ (plot d). We used $k = 1$
    as well in figure (5).
    }
    \label{fig:fig5}
\end{figure}
  So the above eq. dictates that in any of the path $S_D^{(2)}(\Delta \tau)$ takes the following form,
    \begin{eqnarray}
        S_D^{(2)}(\Delta \tau)=N \frac{(\Gamma(D/2))^2}{\Gamma(D-1)}G_{AdS_{2D}}(\Delta \tau). \label{issues}
    \end{eqnarray}
    From eq. \eqref{chhapu} we can then conclude that detector response function for fermions can be related to response function for the scalars,
    \begin{eqnarray}
        \mathcal{J}_{AdS_D}= N \frac{(\Gamma(D/2))^2}{\Gamma(D-1)}
              \mathcal{F}^{(1)}_{\rm AdS_{2D}}, \label{issue}
           \end{eqnarray}
as advertised in theorem 2.
As we explicitly evaluated  $ \mathcal{F}^{(n)}_{\rm AdS_{2D}}(\Delta \tau)$
for any $D$, the fermionic response function can be evaluated just using the above equation just by setting $n=1$.
Following eq. \eqref{path2} it is also evidently clear that 
detector response function for fermions are also path independent just like scalar fields if we consider the uniform acceleration.
In the limit $k\rightarrow 0$, we obtain the exact result of 
the response function related to
Dirac fields in the
flat spacetime. In Figure $5$ (a) and (b) we plotted the fermionnic response function for d=4, with respect to energy and acceleration, respectively. Finally we compared the fermionic response function with the bosonic response function with $n=1$. However as we have seen that detector response is proportional to Bose-Einstein factor in flat spacetime\cite{Nottingham} as well as in AdS for any $D$ $(D\geq2)$, the story with Rindler power spectrum is different. 
 This is due to the definition of the detector response function which is basically dependent upon trace over $4$-point correlator for detector  coupling of the form \eqref{interHam2}. On the other hand Rindler power spectrum is defined as
 trace over $2$-point fermionic correlator.
 Therefore
 Rindler power spectrum of the fermionic field 
maintains a Bose-Einstein factor  for odd dimensions but a Fermi-Dirac factor for even dimensions ($D>2$).\\\\
Until now we have limited our discussion to uniformly accelerated detectors. If we want to broaden the scope of theorem 2 for not only for uniform acceleration but for more general  stationary paths, we require a formalism that will relate the stationary paths in $D$ dimensional AdS spacetime to $2D$ dimensional AdS spacetime. Usually a particle's motion is called stationary if that particle’s world line is the orbit of some time-like Killing vector field. In other words, when geometric properties particle's trajectory are invariant under proper time translation\cite{Letaw}. 
In appendix C we have outlined a new formalism closely related to Frenet-Serret formalism \cite{ Letaw, Iyer} that could be used to construct and categorize stationary trajectories in $D$ dimensional AdS spacetime. In this formalism we embed $D$ dimensional AdS space-time into $D+1$ dimensional flat spacetime and construct the stationary paths that is contained within the intersection of $M$-plane and AdS hypersurface. If a stationary trajectory is contained within the intersection of $M$-plane and $AdS_D$ hyperrsurface, we would say the trajectory is contained in $M$ dimensions. Using this formalism it is possible to classify the stationary paths in AdS spacetime by the minimum number of dimensions that would be necessary to contain the path in the embedding space. Therefore the stationary trajectories of the detector in $AdS_{2D}$ which can be contained at most within $D$-plane could be compared to the stationary trajectories of the detector in $D$ dimension. However  the validity of the equation (80) depends on the validity of the equation (79). The both sides of the equation (79) depend on the conformal invariant in their respective dimension. Therefore, if one stationary trajectory is in $2D$ dimension (\textbf{trajectory B}) and another stationary trajectory in $D$ dimension (\textbf{trajectory A}) have the same conformal invariant $v(\Delta \tau)$, then the equation (79) would be valid for those two particular stationary paths. As a consequence when we take the \textbf{trajectory B} to be in
uniform linear acceleration in 2D dimensions, one can of course
pick \textbf{trajectory A} to have
uniform linear acceleration in D dimensions because both paths will have same conformal invariant in such example. But what if we choose either of the trajectories to be in more general
stationary path\footnote{We thank the referees for prompting the discussion on general stationary paths.}. Will our formula eq (80) still work?
We have formulated a sufficient condition for the two paths-\\ \\ a) to have same functional form of conformal invariant,\\ b)  the minimal containment of these two paths into the planes with same number of dimensions.\\\\
When (a) and (b) both are maintained, one can make sure eq. (80) is  valid for such case.
In the appendix we have shown that for any stationary trajectory in $AdS_D$ the minimal containment dimension and functional form of conformal invariant can uniquely be determined from the squared norm of first $D$ number of  proper derivatives of velocity vectors with respect to propertime. From this conclusion we can deduce the condition (a) and (b) will be maintained if both \textbf{trajectory A} and \textbf{trajectory B} are stationary paths as well as
 have the the same squared norm of first $2D$ number of covariant derivatives of their velocity vectors with respect to propertime. 
 It is of course possible to find paths which are stationary but do not satisfy the criteria to be valid for eq. (80)
 as they do not
  share the same squared norm of first $2D$  derivatives of their velocity vectors.
For those paths we can not use theorem 2 (Eq. 80) to compute the fermionic response function. \\\\

\section{Concluding remarks}
In this article we have analysed detector response function of an accelerated detector
with uniform linearly accelerated timelike path.
For both bosonic and fermionic fields
following the sub-critical and critical paths
accelerated particle detector does not manifest any response. But once we have acceleration greater than the  curvature, we have a non zero response function
while the KMS condition holds at temperature, $ T=\frac{\sqrt{a^2-k^2}}{2\pi}$.
The consequence of introducing non linearity in the interaction Lagrangian between scalar field and detector is also clear from our investigation. While in the  usual linear interaction we have detector response of fermionic signature in odd dimensions it is not in general true for any $n$ in (\ref{kukumia}).
Only for odd $n$ in odd dimensions we will have Fermi-Dirac
contribution in the response function.
 Nonetheless
nonlinearity also alters the shape and characteristics of  the  response function in even dimensions 
as clear from figure (1) - (4). We have also explored the fermionic Unruh effect in AdS for the first time in this article.
There is no distinction noticed for scalar or fermionic response if the detector  follow path \eqref{eq12}
or \eqref{eqnxt}.
Interestingly just like flat spacetime, fermion response function also depends upon Bose Einstein
distribution in AdS spacetime.
The quadratically coupled fermionic
detector's response in AdS spacetime is proportional to that of a detector coupled
linearly to a massless scalar field in $2D$-dimensional AdS spacetime. Same proportionality condition has been derived for Minkowski spacetime in \cite{Nottingham}. 
Although the detector response depends upon curvature, the consequences of theorem $1$ and $2$ applies to AdS as well as Minkowski spacetime\cite{sri, Nottingham}.
\\\\
There are several topics we are focusing immediately including but not limited to-\\\\
(a) We are employing  numerical techniques to study the response function in  $AdS_D$ (odd $D$)
as closing the contour
of \eqref{eqn:detrate}
for such case is not known\footnote{ See page 6 of \cite{Jennings} for further comments.} (figure 1).
Also we are looking for some advanced techniques to solve such Fourier transform analytically. 
Finally we can directly utilize the
obtained
results  in this article to design the Unruh heat engines \cite{ottoEngine,engine2} in AdS.
 The 
present 
literature on Unruh heat engines are 
based upon flat spacetime and deals with linear and quadratic interactions. On the other hand our detector response is for AdS and also is valid for any $n$. We can also reproduce the flat spacetime result for any $n$ in the vanishing curvature limit. 
\\\\\
(b)
Our current study
is focused upon
the vacuum response of a particle detector in Anti-de Sitter spacetime
where we considered non interacting 
 fields (just coupled to background gravity).  In ref. \cite{unruhpath} a path integral formalism was
developed by Unruh
to
show that for a large class of interacting scalar
and spinor field theories in Minkowski spacetime, an accelerated observer sees a
thermal spectrum of particles.
We ought to extend the results for interacting theory in AdS spacetime
using 
path integral formalism .
\\\\(c)
 We also would like to note that using more sophisticated mathematical framework we are working on applying the formalism described in Appendix $C$ to investigate the applicability
 of theorem 2 for stationary paths in other curved spacetimes e.g. de Sitter (dS) spacetime,  Schwarzschild black hole, RN-AdS black hole etc. 
 We have already extended the main statement of theorem 2 for fermions in dS background\cite{shanew}, but to classify the stationary paths 
 in a more complicated background geometries is an interesting  problem to address\cite{Iyer,Letaw,Townsend,Bini,y,shanew}.
\\\\
(d) Finally it has been suggested that dS spacetime can be included in the fundamental landscape of string theory by considering it not as a vacuum but as a coherent state over a Minkowski vacuum \cite{divali1,keshav4}. However this is not the only motivation to consider dS as a coherent state because treating dS as a vacuum imposes quite a lot of problems in the field theory structure. For example-\\\\ (1) fluctuations over a de Sitter
vacuum typically makes the frequencies time-dependent. This makes it almost impossible to construct a well-defined Wilsonian effective action\cite{Weinberg72}.\\\\
(2) As pointed out by Brandenberger and Martin \cite{brandenbarger66}, there are also trans-Planckian issues that do not have simple resolutions.\\\\ (3) The non-closure of the supersymmetry algebra over the de Sitter vacuum means that the zero point energies cannot be easily cancelled or renormalized \cite{Weinberg72}.\\\\ The only way out of these conundrums is to view the de Sitter space as a coherent state over a supersymmetric vacuum. The fact that the vacuum remains supersymmetric means that the cancellation of the zero point energies is guaranteed. However, the supersymmetry is subsequently broken spontaneously by the Glauber-Sudarshan state. The broken supersymmetry is manifest due to the presence of non self-dual fluxes over the compact extra dimensions. Thus taking de Sitter space as a coherent state (but not as a vacua) will naturally cure many of the shortcomings like supersymmetry breaking, finite entropy, cosmological constant problem etc. However it is still not clear how a detector detects a thermal response at Gibbons Hawking temperature, $T_{dS}$ from a coherent state configuration. We are focusing on that at this stage.\\\\
(e) Finally we would like to understand better the strongly coupled field theories using gauge gravity duality dictionary. There should be a simple realization of statistics inversion from the CFT side utilizing  Parikh and   Samantray's approach of Rindler-AdS/CFT 
\cite{rindler}. \\ \\
(f)  In our current article we have analysed a special case of a uniformly  accelerated 
observer coupled to a field in its AdS vacuum.
 We have assumed here that
 the detector interaction time, $\mathcal{T}$ is long enough as well as the process of interaction switch-on and switch-off are
sufficiently slow. Also the back-reaction of the observer on the quantum field remains
small enough to be ignored. In such assumption only super critical acceleration gives non-zero response  of the detector. However, the response function of the form \eqref{eqn:detrate} and \eqref{chhapu} results as we take limit $\mathcal{T}\rightarrow \infty$
in
eq. ($D.1$) of ref.\cite{ottoEngine}. When we take such limit the   sub critical and critical paths result in zero response function\cite{Jennings}.
However following the techniques of ref.\cite{ottoEngine} one can further compute the response function when
interaction time $\mathcal{T}$ is finite.
In practice the interaction time between detector and matter fields  would rather be finite.  
For example to design a realistic  heat engine \footnote{where we put a qubit 
through the quantum equivalent of the Otto cycle and the heat reservoirs are due to the
Unruh effect.}
in 
AdS
spacetime we
need to evaluate the finite time response function\cite{ottoEngine}.  
We have successfully computed the finite time response numerically to construct the Unruh Otto engines and will be reported soon\cite{shanew1}.
\section{Acknowledgement}
The authors would like to thank Keshav Dasgupta, Maxim Emelin, 
Suddhasattwa Brahma, Sebastian Fischetti, Onirban Islam
and Tal Sheaffar for helpful discussions.

\appendix

\section{Fermionic 2-Point correlator}
 We start with \(D\) dimensional AdS spacetime  in Poincare
coordinates.  The Christoffel symbols for such setup-
\begin{eqnarray}
    \Gamma^{\alpha}_{\beta \sigma} &=& \frac{1}{2} g^{a \lambda} \big( \partial_{\beta}g_{\lambda \sigma} + \partial_{\sigma} g_{\beta \lambda} - \partial_{\lambda} g_{\beta \sigma} \big) \nonumber\\
    &=& - g^{\alpha \lambda} \bigg( \frac{\delta^{z}_{\beta}g_{\lambda \sigma}}{z} + \frac{\delta^{z}_{\sigma}g_{\lambda \beta}}{z} - \frac{\delta^{z}_{\lambda}g_{\beta \sigma}}{z} \bigg)\nonumber\\
    &=&\frac{1}{z}\big( g^{\alpha z}g_{\beta \sigma} - \delta^{\alpha}_{\sigma}\delta^{z}_{\beta} - \delta^{\alpha}_{\beta}\delta^{z}_{\sigma}\big).
\end{eqnarray} 
Here, \(\mu, \nu\) run from \(0\) to \(D-1\),
where \(x^0 = t\) and \(x^{D-1} = z\).\\\\
Now the spin connections in eqn. (\ref{spinConn}) become-
\begin{eqnarray}
    \omega^{ab}_{\mu} &=& e^{a \lambda} \bigg( \partial_{\mu}\big(\frac{\delta^b_{\lambda}}{kz}\big)-\Gamma^{\alpha}_{\mu\lambda}e^b_{\alpha} \bigg) \nonumber\\
    &=& e^a_{\beta} g^{\beta \lambda} \bigg( -\frac{\delta^z_{\mu}e^b_{\lambda}}{z}-\frac{e^b_{\alpha}}{z}\big( g^{\alpha z}g_{\mu \lambda}-\delta^{\alpha}_{\lambda}\delta^z_{\mu} - \delta^{\alpha}_{\mu}\delta^z_{\lambda} \big) \bigg)\nonumber\\
    &=& \frac{g^{\beta z}}{z}\big( e^b_{\mu}e^{a}_{\beta}-e^{a}_{\mu}e^{b}_{\beta} \big).
\end{eqnarray}

    Finally, the Dirac operator is defined as
\(\Gamma^{\mu}{D_{\mu}}=\gamma^a e_a^{\mu}( \partial_{\mu} + \frac{1}{2}\omega^{bc}_{\mu}\Omega_{bc})\)
where \(\gamma^{b}\) are a set of \(N \times N\) matrices and
\(\Omega^{bc} = \frac{1}{4}(\gamma^b \gamma^c-\gamma^c \gamma^b)\) \cite{spinorExchange}.
We can simplify \(\omega^{bc}_{\mu}\Omega_{bc}\) to
\(\frac{1}{2}\gamma_{b}\gamma_{c}\omega^{bc}_{\mu}\) by using
anti-symmetric properties of both \(\omega^{bc}_{\mu}\) and
\(\Omega_{bc}\).\\\\ Now in Poincare AdS coordinate the Dirac operator
takes the form, \begin{eqnarray}
   \slashed{D} =  \Gamma^{\mu}{D_{\mu}}&=&\gamma^a e_a^{\mu}\left( \partial_{\mu} + \frac{1}{4}\frac{g^{\beta z}}{z}\big( e^c_{\mu}e^{b}_{\beta}-e^{b}_{\mu}e^{c}_{\beta} \big)\gamma_{b}\gamma_{c}\right) \nonumber \\
    &=& \gamma^a e_a^{\mu}\partial_{\mu} + \frac{g^{\beta z}}{4z} \gamma^a e_a^{\mu} e^c_{\mu}e^{b}_{\beta} \gamma_{b}\gamma_{c} - \frac{g^{\beta z}}{4z} \gamma^a e_a^{\mu} e^{b}_{\mu}e^{c}_{\beta} \gamma_{b}\gamma_{c}\nonumber \\
    &=& \gamma^a e_a^{\mu}\partial_{\mu} + \frac{g^{\beta z}}{4z} e^b_{\beta} \gamma^c  \gamma_{b}\gamma_{c} - \frac{g^{\beta z}}{4z} e^{c}_{\beta} \gamma^b \gamma_{b}\gamma_{c} \label{dirac56}\nonumber \\
    &=& \gamma^a (kz) \delta_a^{\mu}\partial_{\mu} + \frac{g^{\beta z}}{4} e^b_{\beta} \gamma^c \  2 \eta_{bc}- \frac{g^{\beta z}}{4z} e^b_{\beta} \gamma^c  \gamma_{c}\gamma_{b} - \frac{g^{\beta z}}{4z} e^{c}_{\beta} \gamma^b \gamma_{b}\gamma_{c} \label{dirac57}\nonumber \\
    &=& kz \gamma^a  \partial_{a} + \frac{g^{\beta z}}{2z} e^b_{\beta} \gamma_b - \frac{g^{\beta z}}{2z} D e^b_{\beta} \gamma_b\nonumber \\
    &=& kz \gamma^a  \partial_{a} + \frac{g^{\beta z}}{2z} \Gamma_\beta - \frac{g^{\beta z}}{2z} D \Gamma_\beta\nonumber \\
    &=& kz \gamma^a  \partial_{a} - \frac{D-1}{2z} \Gamma^z\nonumber \\
    &=& k\left(z \gamma^a  \partial_{a} - \frac{D-1}{2} \gamma^z \right).
\end{eqnarray} In deriving the above equations we have used some
identities involving gamma matrices, for example,
\(\gamma_b\gamma_c = (2\eta_{bc}-\gamma_c\gamma_b)\) and
\(\gamma^a\gamma_a = D\mathbb{I}_{N\times N}\).\\\\
We are going to use the following representation for
Gamma  matrices: \begin{equation}
\gamma ^{0}=i\left[
\begin{array}{cc}
\mathbf{0}_{(N/2)\times(N/2)} & -\mathbb{I}_{(N/2)\times(N/2)} \\
\mathbb{I}_{(N/2)\times(N/2)} & \mathbf{0}_{(N/2)\times(N/2)}%
\end{array}
\right] ,\;\gamma^{a}=i\left[
\begin{array}{cc}
-\sigma ^{a} & \mathbf{0}_{(N/2)\times(N/2)} \\
\mathbf{0}_{(N/2)\times(N/2)} & \sigma ^{a}
\end{array}
\right] ,  \label{gamflat}
\end{equation} 
with \(a=1,\ldots ,D-1\). 
Here \(\mathbb{I}_{(N/2)\times(N/2)}\) is
an identity matrix of dimension
\({(N/2)\times(N/2)}\). The definition of $N$
can be found in \eqref{chandumama}.
Also \( \mathbf{0}_{(N/2)\times(N/2)}\) is a matrix of dimension \({(N/2)\times(N/2)}\) with zero entries in all of it components.
We also have,
\begin{eqnarray}
    \sigma ^{a}\sigma ^{b}+\sigma ^{b}\sigma^{a} &=& 2\delta^{ab} \mathbb{I}_{(N/2)\times(N/2)}, \nonumber \\
    \sigma^{a \dagger} &=& \sigma^{a}. \nonumber
\end{eqnarray}
Moreover we take explicit form of $\sigma^{z}$ and $\sigma^l$ (\(l = 1, \ldots , D-2\)) in the following form
\begin{eqnarray}
    \sigma ^{z}&=&\left[
    \begin{array}{cc}
\mathbb{I}_{(N/4)\times(N/4)} & \mathbf{0}_{(N/4)\times(N/4)} \\
\mathbf{0}_{(N/4)\times(N/4)} & -\mathbb{I}_{(N/4)\times(N/4)}%
\end{array}%
\right] ,  \label{sigD} \\
\sigma^{l}&=& \left[
\begin{array}{cc}
\mathbf{0}_{(N/4)\times(N/4)} & b^{l} \\
c^{l} & \mathbf{0}_{(N/4)\times(N/4)}%
\end{array}%
\right].
\end{eqnarray}
with \(b^{l}c^{l}=c^{l}b^{l}=\mathbb{I}_{(N/4)\times(N/4)}\) and
\(b^{l}c^{k}=-b^{k}c^{l}\),\(\;c^{l}b^{k}=-c^{k}b^{l}\),
for\(\;l\neq k\).\\\\
Starting with massive Dirac equation,
\begin{equation}
i\slashed{D}\psi -m\psi =0\ .\label{Direq}
\end{equation}\\
Let us first start with (\ref{Direq}) positive energy solutions. These solutions are proportional to 
\(e^{i\mathbf{px}-i\omega t}\), where
\(\mathbf{x}=(x^{1},\ldots ,x^{D-2})\) and
\(\mathbf{p}=(p_{1},\ldots ,p_{D-2})\), \(\mathbf{px}=p_{l}x^{l}\), and
the summation runs over \(l=1,\ldots ,D-2\).
Decomposing the positive energy modes into upper and lower components, \begin{equation}
\psi^{(+)} =\left[
\begin{array}{c}
\psi _{+}(z) \\
\psi _{-}(z)%
\end{array}%
\right] e^{i\mathbf{px}-i\omega t}. \label{decomp1}
\end{equation}
The Dirac equation is then reduced to subsequent form 
for positive energy modes,
\begin{equation}
\left( \sigma ^{z}\left(\partial _{z}-\frac{D-1}{2z}\right) +ip_{l}\sigma^{l}\mp \frac{m}{kz}\right) \psi _{\pm }-i\omega \psi _{\mp }=0.
\label{psipm}
\end{equation}
Using equation (\ref{psipm}) we can deduce two different second order differential equations
for the upper and lower components: 
\begin{equation}
\left( z^{2}\partial _{z}^{2}-(D-1)z\partial _{z}+\lambda ^{2}
z^{2}+\frac{(D-1)^{2}}{4}+\frac{D-1}{2}-\frac{m^{2}}{k^{2}}\pm \sigma ^{z}\frac{m}{k}\right) \psi _{\pm }=0,
\label{Eqpsipm}
\end{equation} where \(\lambda ^{2}=\omega ^{2}-p^{2}\). Making the
following
substitution, \begin{equation}
\psi _{\pm }(z)=z^{D/2}\chi _{\pm }(z),  \label{xipm}
\end{equation} equation (\ref{Eqpsipm}) is reduced to the following form, \begin{equation}
\left( z^{2}\partial _{z}^{2}+z\partial _{z}+\lambda ^{2}z^{2}\pm \sigma
^{z}\frac{m}{k}-\frac{m^{2}}{k^{2}}-\frac{1}{4}\right) \chi _{\pm }=0.  \label{Eqxipm}
\end{equation}
Using equation (\ref{sigD}) we further decompose \(\chi _{\pm }(z)\) into upper and
lower components, \begin{equation}
\chi _{\pm }(z)=\left[
\begin{array}{c}
\varphi _{\pm \uparrow }(z) \\
\varphi _{\pm \downarrow }(z)%
\end{array}%
\right] .  \label{decompxi}
\end{equation}
The solutions for these components directly follow from 
(\ref{Eqpsipm}) \cite{elizalde2013fermionic,
bellucci2017fermionic}, \begin{eqnarray}
\varphi _{\pm \uparrow } &=&C_{\pm \uparrow }^{(J)}J_{m/k\mp 1/2}(\lambda
z)+C_{\pm \uparrow }^{(Y)}Y_{m/k\mp 1/2}(\lambda z),  \notag \\
\varphi _{\pm \downarrow } &=&C_{\pm \downarrow }^{(J)}J_{m/k\pm 1/2}(\lambda
z)+C_{\pm \downarrow }^{(Y)}Y_{m/k\pm 1/2}(\lambda z).  \label{phipm}
\end{eqnarray} where \(J_{\nu }(x)\) and \(Y_{\nu }(x)\) are Bessel
and Neumann functions, respectively. Now we are going to put the following boundary condition over $\psi(z)$.
\begin{eqnarray}
    \lim_{z \to 0} z^{-D/2}  \ \psi(z) = 0 \label{bdcnd}.
\end{eqnarray}
This mimics the same boundary condition in \eqref{traces}.
This will make the coefficient of the Neumann function vanish i.e. \(C_{+\uparrow }^{(Y)} = C_{-\uparrow }^{(Y)} = C_{+\downarrow }^{(Y)} = C_{+\downarrow }^{(Y)} = 0\). Our final goal is to find the 2 point function for massless case. When $m \to 0$ the solution of equation (\ref{Eqxipm}) only has two solutions $J_{\pm 1/2}(z)$, so discarding the Neumann functions will not affect the final result. Choice of this type of boundary conditions has been discussed at length in \cite{elizalde2013fermionic,bellucci2017fermionic}. 
From the equation (\ref{psipm}) with the upper
sign we find out, 
\begin{eqnarray}
&& \omega C_{-\uparrow }^{(J)} =p_{l}b^{l}C_{+\downarrow }^{(J)}+i\lambda
C_{+\uparrow }^{(J)} \label{Cpmu},\\
&& \omega C_{-\downarrow }^{(J)} =p_{l}c^{l}C_{+\uparrow }^{(J)}+i\lambda
C_{+\downarrow }^{(J)} \label{Cpmd}.
\end{eqnarray}

Now introducing the notations to write the solutions in a compact form, 
\begin{eqnarray}
C_{+}^{(J)} &=& \left[
\begin{array}{c}
C_{+\uparrow }^{(J)} \\
C_{+\downarrow }^{(J)}%
\end{array}%
\right] , \label{C+} \\
\widehat{J}_{\pm }(z) &=&\left[
\begin{array}{cc}
\mathbf{J}_{m/k\pm 1/2}(z) & \mathbf{0} \\
\mathbf{0}  & \mathbf{J}_{m/k\mp 1/2}(z) %
\end{array}
\right] .  \label{Zipm}
\end{eqnarray} 
Here we used $\mathbf{0} = \mathbf{0}_{(N/4)\times(N/4)}$, $\widehat{0} = \mathbf{0}_{(N/2)\times(N/2)}$ and $\mathbf{J}_{m/k\pm 1/2}(z) = J_{m/k\pm 1/2}(z)\times\mathbb{I}_{(N/4)\times(N/4)}$.
In the subsequent calculations to avoid clutter we will not explicitly write the identity matrices e.g. we will use \(i\lambda\) instead of \(i\lambda \times \mathbb{I}_{(N/2)\times(N/2)}\). Finally, the solutions obeying
the boundary condition at \(z=0\) can be expressed in the form
\begin{equation}
\psi^{(+)} =z^{D/2}e^{i\mathbf{px}-i\omega t}\left[
\begin{array}{c}
\widehat{J}_{-}(\lambda z)C_{+}^{(J)} \\
\omega ^{-1}\widehat{J}_{+}( \lambda z)\left( i\lambda
+p_{l}\sigma ^{l}\right) C_{+}^{(J)}%
\end{array}%
\right] .  \label{psi3}
\end{equation}

An additional quantum number apart from $\lambda$ and  
\(\mathbf{p}\)
is still needed to specify all the solutions.
So we need to fix the  spinor $C_{+}^{(J)}$. Here we take orthonormal basis for spinors by choosing 
\(C_{+}^{(J)}=C_{\beta }^{(+)}w^{(\sigma )}\), where \(C_{\beta }^{(+)}\) is a normalization constant and \(w^{(\sigma )}\), \(\sigma =\)
\(1,\ldots ,N/2\), are one-column matrices of \(N/2\) rows, with elements \(w_{l}^{(\sigma )}=\delta _{l\sigma }\). Combining with the negative energy solutions this set $\beta = (\mathbf{p}, \lambda, \sigma)$ will form a complete set of quantum numbers. As a result, the positive-energy mode functions obeying the boundary condition at \(z=0\) takes  the following form,
\begin{equation}
\psi _{\beta }^{(+)}=C_{\beta }^{(+)}z^{{D}/{2}}e^{i\mathbf{px}%
-i\omega t}\left[
\begin{array}{c}
\widehat{J}_{-}( \lambda z)w^{(\sigma )} \\
\frac{1}{\omega }\widehat{J}_{+}( \lambda z)\left( i\lambda
+p_{l}\sigma^{l}\right) w^{(\sigma )}%
\end{array}
\right] .  \label{psi+}
\end{equation}

The coefficient \(C_{\beta }^{(+)}\) in (\ref{psi+}) is set on from the normalization condition (using inner product defined over constant time hypersurface) \cite{sahalect}
\begin{equation}
<\psi^{(+)}_{\beta}, \psi^{(+)}_{\beta'}> = \int d^{D-2}x\int_{0}^{\infty}dz\,\sqrt{\frac{|g|}{g^{00}}}\psi _{\beta
}^{(+)\dagger}\psi _{\beta ^{\prime }}^{(+)}=\delta (\mathbf{p-p}^{\prime })\delta
_{\sigma \sigma ^{\prime }}\delta({\lambda - \lambda^{\prime })},  \label{NormCond}
\end{equation} where \(|g|=1/{(kz)}^{2D}\) is the determinant of the metric.  Also,\\
\begin{eqnarray}
    &&\int d^{D-2}x\int_{0}^{\infty}dz\,\sqrt{\frac{|g|}{g^{00}}}\psi _{\beta
}^{(+)\dagger}\psi _{\beta ^{\prime }}^{(+)} \nonumber \\
&=& |C_{\beta }^{(+)}|^{2} \int d^{D-1}x\int_{0}^{\infty}dz (\frac{1}{kz})^{D-1} z^{D} e^{i\mathbf{(p - p')x}} w^{(\sigma) \dagger} \bigg( \widehat{J}_{-}( \lambda z)\widehat{J}_{-}( \lambda' z) +  \nonumber
\\ & &  \ \ \ \ \ \ \ \ \ \ \ \ \ \ \ \ \ \ \ \ \ \ \ \ \ \ \ \ \ \ \ \ \ \ \ \ \ \ \ \ \ \ \ \ \ \ \ \ \ \ \ \ \ \ \ \ \ \ \ \ \ \ \ \ \ \ \ \ \ \ \ \ \ \ \ \ \ \ \ \   \frac{(-i\lambda + p_l \sigma^l)}{\omega} (\widehat{J}_{+}( \lambda z) \widehat{J}_{+}( \lambda' z)) \frac{(i\lambda + p_q \sigma^q)}{\omega} \bigg)  w^{(\sigma')} \nonumber
\\
&=& |C_{\beta }^{(+)}|^{2} \int d^{D-2}x\int_{0}^{\infty}dz (\frac{1}{k})^{D-1} z e^{i\mathbf{(p - p')x}} w^{(\sigma) \dagger} \bigg( \widehat{J}_{-}( \lambda z)\widehat{J}_{-}( \lambda' z) + \nonumber  \\
& &  \ \ \ \ \ \ \ \ \ \ \ \ \ \ \ \ \ \ \ \ \ \ \ \ \ \ \ \ \ \ \ \ \ \ \ \ \ \ \ \ \ \ \ \ \ \ \ \ \ \ \ \ \ \ \ \ \ \ \ \ \ \ \ \ \ \ \ \ \ \ \ \ \ \ \ \ \ \ \ \  
\frac{\lambda^2}{\omega^2} \widehat{J}_{+}( \lambda z)\widehat{J}_{+}( \lambda' z) + \widehat{J}_{-}( \lambda z)\widehat{J}_{-}( \lambda' z) \frac{p_l p_q \sigma^l \sigma_q}{\omega^2} \bigg)  w^{(\sigma')} \nonumber
\\
&=& 
|C_{\beta }^{(+)}|^{2} \int d^{D-2}x\int_{0}^{\infty}dz (\frac{1}{k})^{D-1} z e^{i\mathbf{(p - p')x}} w^{(\sigma) \dagger} \bigg( \frac{\lambda^2}{\omega^2} \widehat{J}_{+}( \lambda z)\widehat{J}_{+}( \lambda' z)  +
\widehat{J}_{-}( \lambda z)\widehat{J}_{-}( \lambda' z) \frac{\omega^2+\mathbf{p}^2}{\omega^2} \bigg)  w^{(\sigma')} \nonumber
\\
&=& \frac{(2\pi)^{D-2}|C_{\beta }^{(+)}|^{2}}{k^{D-1}} w^{(\sigma) \dagger} \bigg( \int \frac{d^{D-2}x}{(2\pi)^{D-2}} e^{i\mathbf{(p - p')x}} \int_{0}^{\infty}dz \big(  z \frac{\lambda^2}{\omega^2} \widehat{J}_{+}( \lambda z)\widehat{J}_{+}( \lambda' z) + \nonumber \\
& &  \ \ \ \ \ \ \ \ \ \ \ \ \ \ \ \ \ \ \ \ \ \ \ \ \ \ \ \ \ \ \ \ \ \ \ \ \ \ \ \ \ \ \ \ \ \ \ \ \ \ \ \ \ \ \ \ \ \ \ \ \ \ \ \ \ \ \ \ \ \ \ \ \ \ \ \ \ \ \ \  \ \ \ \ \ \ \ \ \ \ \ \ \ \ \ \ \ \ \ \ \ \ \ \ \ \ \ \ 
 z \widehat{J}_{-}( \lambda z)\widehat{J}_{-}( \lambda' z) \frac{\omega^2+\mathbf{p}^2}{\omega^2}  \big) \bigg)  w^{(\sigma')} \nonumber \\
 &=& \frac{(2\pi)^{D-2}|C_{\beta }^{(+)}|^{2}}{k^{D-1}} w^{(\sigma) \dagger} \bigg( \delta(\mathbf{p- p'}) \frac{\delta(\lambda - \lambda')}{\lambda} \big(\frac{\lambda^2+\omega^2+\mathbf{p}^2}{\omega^2}\big)\bigg)  w^{(\sigma')} \nonumber \\
 &=& \frac{2(2\pi)^{D-2}|C_{\beta }^{(+)}|^{2}}{\lambda k^{D-1}}  \delta(\mathbf{p- p'}) {\delta(\lambda - \lambda')}   \delta_{\sigma \sigma'}. \nonumber
\end{eqnarray}\\
Thus the normalization constant is,
\begin{equation}
|C_{\beta }^{(+)}|^{2}=\frac{\lambda k^{D-1}}{2 (2\pi)^{D-2}}.  \label{Cbet}
\end{equation} 

The negative-energy mode functions can be obtained in a similar way. We can start from the ansatz for negative energy spinor as
\begin{equation}
\psi^{(-)} =\left[
\begin{array}{c}
\psi _{+}(z) \\
\psi _{-}(z)%
\end{array}%
\right] e^{i\mathbf{px}+i\omega t}.  \label{decomp2}
\end{equation}
Just like for positive energy solutions we will put equation (\ref{bdcnd}) as boundary condition. However, we can find the relation between $C^{(J)}_{+ \uparrow \downarrow}$ and $C^{(J)}_{- \uparrow \downarrow}$ by putting $\omega \to -\omega$ in equations (\ref{Cpmd}) and (\ref{Cpmu}). Then we can represent $C^{(J)}_{+ \uparrow \downarrow}$ in terms of $C^{(J)}_{- \uparrow \downarrow}$ in the following manner.
\begin{eqnarray}
    \left[ \begin{array}{c}
          C^{(J)}_{+ \uparrow}  \\
         C^{(J)}_{+ \downarrow} 
    \end{array} \right] &=& \frac{1}{\omega} (i\lambda - p_q \sigma^q)\left[ \begin{array}{c}
         C^{(J)}_{- \uparrow}  \\
         C^{(J)}_{- \downarrow} 
    \end{array} \right].
\end{eqnarray}

 Thus the negative energy solutions have the form: \begin{equation}
\psi _{\beta }^{(-)}=C_{\beta }^{(-)}z^{D/2}e^{i\mathbf{px}%
+i\omega t}\left[
\begin{array}{c}
\frac{1}{\omega }\widehat{J}_{-}( \lambda z)\left( i\lambda
-p_{l}\sigma ^{l}\right) w^{(\sigma )} \\
\widehat{J}_{+}( \lambda z) w^{(\sigma )}%
\end{array}%
\right] ,  \label{psi-}
\end{equation} where \(|C_{\beta }^{(-)}|^{2} = |C_{\beta }^{(+)}|^{2}\). 
We can further make sure that these negative energy modes are orthrogonal to the previous $\psi^{(+)}_{\beta}$ by taking the inner product between them. 
\begin{eqnarray}
    <\psi^{(+)}_{\beta}, \psi^{(-)}_{\beta'}> &= &\int d^{D-2}x\int_{0}^{\infty}dz\,\sqrt{\frac{|g|}{g^{00}}}\psi _{\beta
}^{(+)\dagger}\psi _{\beta ^{\prime }}^{(-)} \nonumber \\
&=& 
|C_{\beta }^{(+)}|^{2} \int d^{D-1}x\int_{0}^{\infty}dz (\frac{1}{kz})^{D-1} z^{D} e^{i\mathbf{(p - p')x}} w^{(\sigma) \dagger} \bigg( \widehat{J}_{-}( \lambda z)\widehat{J}_{-}( \lambda' z) \frac{(i\lambda - p_q \sigma^q)}{\omega} +  \nonumber
\\ & &  \ \ \ \ \ \ \ \ \ \ \ \ \ \ \ \ \ \ \ \ \ \ \ \ \ \ \ \ \ \ \ \ \ \ \ \ \ \ \ \ \ \ \ \ \ \ \ \ \ \ \ \ \ \ \ \ \ \ \ \ \ \ \ \ \ \ \ \ \ \ \ \ \ \ \ \ \ \ \ \   \frac{(-i\lambda + p_l \sigma^l)}{\omega} (\widehat{J}_{+}( \lambda z) \widehat{J}_{+}( \lambda' z))  \bigg)  w^{(\sigma')} \nonumber
\\
&=& |C_{\beta }^{(+)}|^{2} \int d^{D-1}x (\frac{1}{k})^{D-1}  e^{i\mathbf{(p - p')x}} w^{(\sigma) \dagger} \bigg( \frac{\delta(\lambda - \lambda')}{\lambda} \frac{(i\lambda - p_q \sigma^q)}{\omega} +  \nonumber
\\ & &  \ \ \ \ \ \ \ \ \ \ \ \ \ \ \ \ \ \ \ \ \ \ \ \ \ \ \ \ \ \ \ \ \ \ \ \ \ \ \ \ \ \ \ \ \ \ \ \ \ \ \ \ \ \ \ \ \ \ \ \ \ \ \ \ \ \ \ \ \ \ \ \ \ \ \ \ \ \ \ \   \frac{(-i\lambda + p_l \sigma^l)}{\omega} \frac{\delta(\lambda - \lambda')}{\lambda}  \bigg)  w^{(\sigma')} \nonumber\\
&=& 0.
\end{eqnarray}
Hence we have managed to find a complete set of solutions for the Dirac equation (\ref{Direq}). Now we can write artibitary spinor solution $\Psi(x)$ in the operator form.
\begin{eqnarray}
    \Psi(x) &=& \sum_{\sigma = 1}^{N/2} \int d\mathbf{p} \int_{0}^{\infty} d\lambda \  \bigg( b_{\sigma}(\mathbf{p}, \lambda) \psi^{(+)}_{\sigma}(\mathbf{p}, \lambda, x) + d^{\dagger}_{\sigma}(\mathbf{p}, \lambda) \psi^{(-)}_{\sigma}(\mathbf{p}, \lambda, x)  \bigg) \\
    \overline {\Psi}(x) &=& \sum_{\sigma = 1}^{N/2} \int d\mathbf{p} \int_{0}^{\infty} d\lambda \  \bigg( b^{\dagger}_{\sigma}(\mathbf{p}, \lambda) \overline{{\psi}_{\sigma}^{(+)}}(\mathbf{p}, \lambda, x) + d_{\sigma}(\mathbf{p}, \lambda) \overline{\psi_{\sigma}^{(-)}}(\mathbf{p}, \lambda, x)  \bigg) ,
\end{eqnarray}
where 
\begin{eqnarray}
    b_{\sigma}(\mathbf{p}, \lambda) \ket{0} &=& d_{\sigma}(\mathbf{p}, \lambda) \ket{0} = 0,\\
    \overline{\psi} &=& \psi^{\dagger} \gamma^{0},\\
    \{b_{\sigma}(\mathbf{p}, \lambda), b^{\dagger}_{\sigma'}(\mathbf{p'}, \lambda') \} &=& \delta({\mathbf{p - p'}}) \delta({\lambda - \lambda'}) \delta_{\sigma \sigma'}, \\
    \{d_{\sigma}(\mathbf{p}, \lambda), d^{\dagger}_{\sigma'}(\mathbf{p'}, \lambda') \} &=& \delta({\mathbf{p - p'}}) \delta({\lambda - \lambda'}) \delta_{\sigma \sigma'}.
\end{eqnarray}
Now we can define the fermion propagator as,
\begin{eqnarray}
  S^{+}(x,x')  &=& \bra{0} \Psi(x) \overline{\Psi}(x') \ket{0} \nonumber\\
    &=& \sum_{\sigma \sigma'} \int d\mathbf{p} d\mathbf{p'} \int_{0}^{\infty} d\lambda d\lambda' \psi^{(+)}_{\sigma}(\mathbf{p}, \lambda, x)   \bra{0} b_{\sigma}(\mathbf{p}, \lambda) b^{\dagger}_{\sigma'}(\mathbf{p'}, \lambda')\ket{0} \overline{{\psi}_{\sigma'}^{(+)}}(\mathbf{p'}, \lambda', x') \nonumber \\
    &=& \sum_{\sigma \sigma'} \int d\mathbf{p} d\mathbf{p'} \int_{0}^{\infty} d\lambda d\lambda' \psi^{(+)}_{\sigma}(\mathbf{p}, \lambda, x)    \delta({\mathbf{p - p'}}) \delta({\lambda - \lambda'}) \delta_{\sigma \sigma'}
    \overline{{\psi}_{\sigma'}^{(+)}}(\mathbf{p'}, \lambda', x') \nonumber \\
    &=& \sum_{\sigma }  \int d\mathbf{p} \int_{0}^{\infty} d\lambda \  \psi^{(+)}_{\sigma}(\mathbf{p}, \lambda, x)    
    \overline{{\psi}_{\sigma}^{(+)}}(\mathbf{p}, \lambda, x') \label{Splussss}.
\end{eqnarray}
If we put equation (\ref{psi+}) into (\ref{Splussss}), we obtain-
\begin{eqnarray}
    & & S^{+}(x,x') \nonumber\\
    &=& \sum_{\sigma }  \int d\mathbf{p} \int_{0}^{\infty} d\lambda \ |C_{\beta }^{(+)}|^2 {(zz')}^{{D}/{2}}e^{i\mathbf{p(x-x')}-i\omega (t-t')}\left[
    \begin{array}{c}
    \widehat{J}_{-}( \lambda z)w^{(\sigma )} \\
    \frac{1}{\omega }\widehat{J}_{+}( \lambda z)\left( i\lambda
    +p_{l}\sigma^{l}\right) w^{(\sigma )}%
    \end{array} \right] \times \nonumber \\
    & &  \ \ \ \ \ \ \ \ \ \ \ \ \ \ \ \ \ \ \ \ \ \ \ \ \ \ \ \ \ \ \ \ \ \ \ \ \ \ \ \ \ \ \ \ \ \ \ \ \ \ \ \ \ \ \ \ \ \ \ \ \ \ \ \ \ \ \ \ \ \ \ \ \ \ \ \ \ \ \ \ 
    (-i) \bigg[ w^{(\sigma ) \dagger} \frac{\left( i\lambda
    -p_{q}\sigma^{q}\right)}{\omega} \widehat{J}_{+}(\lambda z')  \ \ w^{(\sigma ) \dagger} \widehat{J}_{-}( \lambda z') \bigg] \nonumber\\
     &=& \int d\mathbf{p} \int_{0}^{\infty} d\lambda \ |C_{\beta }^{(+)}|^2 {(zz')}^{{D}/{2}}e^{i\mathbf{p(x-x')}-i\omega (t-t')} (-i) \times \nonumber  \\
     & &  \left[
     \begin{array}{cc}
        \widehat{J}_{-}( \lambda z) \left( \sum_{\sigma } w^{(\sigma )}   w^{(\sigma ) \dagger} \right) \frac{\left( i\lambda
    -p_{l}\sigma^{l}\right)}{\omega} \widehat{J}_{+}(\lambda z')  & \widehat{J}_{-}( \lambda z)\left( \sum_{\sigma } w^{(\sigma )}   w^{(\sigma ) \dagger} \right) \widehat{J}_{-}( \lambda z') \\
       \frac{1}{\omega }\widehat{J}_{+}( \lambda z)\left( i\lambda
    +p_{l}\sigma^{l}\right) \left( \sum_{\sigma } w^{(\sigma )}   w^{(\sigma ) \dagger} \right) \frac{\left( i\lambda
    -p_{q}\sigma^{q}\right)}{\omega} \widehat{J}_{+}(\lambda z')  & \frac{1}{\omega }\widehat{J}_{+}( \lambda z)\left( i\lambda
    +p_{q}\sigma^{q}\right) \left( \sum_{\sigma } w^{(\sigma )}   w^{(\sigma ) \dagger} \right) \widehat{J}_{-}( \lambda z') 
     \end{array}
     \right] \nonumber \\
      &=& \int d\mathbf{p} \int_{0}^{\infty} d\lambda \ |C_{\beta }^{(+)}|^2 {(zz')}^{{D}/{2}}e^{i\mathbf{p(x-x')}-i\omega (t-t')} (-i) \times \nonumber  \\
     & &  \left[
     \begin{array}{cc}
       \frac{i\lambda}{\omega} \widehat{J}_{-}( \lambda z)\widehat{J}_{+}(\lambda z')  - \frac{\left(    p_{l}\sigma^{l}\right)}{\omega} \widehat{J}_{+}(\lambda z)   \widehat{J}_{+}(\lambda z')  
    &
    \widehat{J}_{-}( \lambda z) \widehat{J}_{-}( \lambda z') \\
       \frac{1}{\omega }\widehat{J}_{+}( \lambda z) \frac{\left( -\omega^2\right)}{\omega} \widehat{J}_{+}(\lambda z')  
       &
       \frac{i\lambda}{\omega}\widehat{J}_{+}( \lambda z) \widehat{J}_{-}( \lambda z')  + \frac{\left( p_{l}\sigma^{l}\right)}{\omega} \widehat{J}_{-}( \lambda z)  \widehat{J}_{-}( \lambda z') \nonumber
     \end{array}
     \right] \nonumber \\
     &=& \int d\mathbf{p} \int_{0}^{\infty} d\lambda \ \frac{|C_{\beta }^{(+)}|^2}{\omega} {(zz')}^{{D}/{2}}e^{i\mathbf{p(x-x')}-i\omega (t-t')} \ \   \times \nonumber  \\
     & & \left( (-i) \left[
     \begin{array}{cc}
       {i\lambda} \widehat{J}_{-}( \lambda z)\widehat{J}_{+}(\lambda z') 
    & \widehat{0} \\
    \widehat{0} &
       i \lambda \widehat{J}_{+}( \lambda z) \widehat{J}_{-}( \lambda z')   
     \end{array} 
     \right] + (\omega \gamma^0 - p_l \gamma^l)
     \left[
     \begin{array}{cc}
         \widehat{J}_{+}(\lambda z)   \widehat{J}_{+}(\lambda z')  
    & \widehat{0} \\
      \widehat{0}   &
        \widehat{J}_{-}( \lambda z)  \widehat{J}_{-}( \lambda z') \nonumber
     \end{array}
     \right]
     \right) \nonumber\\
     &=& \int d\mathbf{p} \int_{0}^{\infty} d\lambda \ \frac{|C_{\beta }^{(+)}|^2}{\omega} {(zz')}^{{D}/{2}}e^{i\mathbf{p(x-x')}-i\omega (t-t')} \ \    \bigg( (\omega \gamma^0 - p_l \gamma^l)
     \left[
     \begin{array}{cc}
         \widehat{J}_{+}(\lambda z)   \widehat{J}_{+}(\lambda z')  
    & \widehat{0} \\
      \widehat{0}   &
        \widehat{J}_{-}( \lambda z)  \widehat{J}_{-}( \lambda z') 
     \end{array}
     \right]  \nonumber  \\
     & &  \ \ \ \ \ \ \ \ \ \ \ \ \ \ \ \ \ \ \ \ \ \ \ \ \ \ \ \ \ \ \ \ \ \ \ \ \ \ \ \ \ \ \ \ \ \ \ \ \ \ \ \ \ \ \ \ \ \ \ \ \ \ \ \ \ 
      - i \left[
     \begin{array}{cc}
       {i\lambda} \widehat{J}_{-}( \lambda z)\widehat{J}_{+}(\lambda z') 
    & \widehat{0} \\
    \widehat{0} &
       i \lambda \widehat{J}_{+}( \lambda z) \widehat{J}_{-}( \lambda z')   
     \end{array} 
     \right] 
     \bigg) \label{j+j-}.
\end{eqnarray}
In equation (\ref{j+j-}), for the second term we can use the following identity to simplify it more,
\begin{eqnarray}
    \lambda \widehat{J}_{\pm}( \lambda z) &=& \left[\begin{array}{cc}
         \lambda \mathbf{J}_{m/k \pm 1/2} (\lambda z) & \mathbf{0}  \\
         \mathbf{0} & \lambda \mathbf{J}_{m/k \mp 1/2} (\lambda z)  
    \end{array} \right] \nonumber \\
    &=& \left[ \begin{array}{cc}
         \mp \partial_z \mathbf{J}_{\frac{m}{k} \mp \frac{1}{2}} (\lambda z) + \left(\frac{m}{k} \mp \frac{1}{2}\right) {\mathbf{J}_{m/k \mp 1/2} (\lambda z)}/{z} & \mathbf{0}  \\
         \mathbf{0} & \pm \partial_z \mathbf{J}_{\frac{m}{k} \pm \frac{1}{2}} (\lambda z) + \left(\frac{m}{k} \pm \frac{1}{2}\right) {\mathbf{J}_{m/k \pm 1/2} (\lambda z)}/{z} 
    \end{array} \right] \nonumber \\
    &=& (\partial_z - \frac{1}{2z}) \left[\begin{array}{cc}
         \mp \mathbf{J}_{\frac{m}{k} \mp \frac{1}{2}} (\lambda z) & \mathbf{0}  \\
         \mathbf{0} & \pm \mathbf{J}_{\frac{m}{k} \pm \frac{1}{2}} (\lambda z)  
    \end{array} \right] + \frac{m}{kz}\left[\begin{array}{cc}
          \mathbf{J}_{\frac{m}{k} \mp \frac{1}{2}} (\lambda z) & \mathbf{0}  \\
         \mathbf{0} &  \mathbf{J}_{\frac{m}{k} \pm \frac{1}{2}} (\lambda z)  
    \end{array} \right] . \label{lamjpm}
\end{eqnarray}\\
Finally, we can substitute $\lambda \widehat{J}_{\pm}( \lambda z)$ in the second term of the equation (\ref{j+j-}),
\begin{eqnarray}
    && \left[
     \begin{array}{cc}
       {\lambda} \widehat{J}_{-}( \lambda z)\widehat{J}_{+}(\lambda z') 
    & \widehat{0} \\
    \widehat{0} &
       \lambda \widehat{J}_{+}( \lambda z) \widehat{J}_{-}( \lambda z')   
     \end{array} 
     \right] = \left[
     \begin{array}{cc}
       {\lambda} \widehat{J}_{-}( \lambda z)
    & \widehat{0} \\
    \widehat{0} &
       \lambda \widehat{J}_{+}( \lambda z)  
     \end{array} 
     \right] \left[
     \begin{array}{cc}
       \widehat{J}_{+}(\lambda z') 
    & \widehat{0} \\
    \widehat{0} &
     \widehat{J}_{-}( \lambda z')   
     \end{array} 
     \right] \nonumber \\
    &=& \left((\partial_z + \frac{1}{2z}) \left[\begin{array}{cccc}
         \mathbf{J}_{\frac{m}{k} + \frac{1}{2}}(\lambda z) & \mathbf{0} & \mathbf{0} & \mathbf{0}  \\
         \mathbf{0} & -\mathbf{J}_{\frac{m}{k} - \frac{1}{2}}(\lambda z) & \mathbf{0} & \mathbf{0}  \\
         \mathbf{0} & \mathbf{0} & -\mathbf{J}_{\frac{m}{k} - \frac{1}{2}}(\lambda z) & \mathbf{0}  \\
         \mathbf{0} & \mathbf{0} & \mathbf{0} & \mathbf{J}_{\frac{m}{k} + \frac{1}{2}} (\lambda z)  
    \end{array} \right] + \frac{m}{kz} \left[
     \begin{array}{cc}
       \widehat{J}_{+}(\lambda z') 
    & \widehat{0} \\
    \widehat{0} &
     \widehat{J}_{-}( \lambda z')   
     \end{array} 
     \right] \right) \nonumber \\
     & & \ \ \ \ \ \ \ \ \ \ \ \ \ \ \ \ \ \ \ \ \ \ \ \ \ \ \ \ \ \ \ \ \ \ \ \ \ \ \ \ \ \ \ \ \ \ \ \ \ \ \ \ \ \ \ \ \ \ \ \ \ \ \ \ \ \ \ \ \ \ \ \ \ \ \ \ \ \ \ \ \ \ \ \ \ \ \ \ \ \ \ \ \ \ \ \ \ \ \ \ \ \ \ \ \ \ \ \ \ \ \ \ \ \ \ \ \ \ \ \
    \times \left[
     \begin{array}{cc}
       \widehat{J}_{+}(\lambda z') 
    & \widehat{0} \\
    \widehat{0} &
     \widehat{J}_{-}( \lambda z')   
     \end{array} 
     \right] \nonumber \\
     &=& \left((i\gamma^z) (\partial_z + \frac{1}{2z}) + \frac{m}{kz} \right) \left[
     \begin{array}{cc}
         \widehat{J}_{+}(\lambda z)   \widehat{J}_{+}(\lambda z')  
    & \widehat{0} \\
      \widehat{0}   &
        \widehat{J}_{-}( \lambda z)  \widehat{J}_{-}( \lambda z') 
     \end{array}
     \right] \label{lamjj}.
\end{eqnarray}
Putting the formula from equation (\ref{lamjj}) to  (\ref{j+j-}), we have-
\begin{eqnarray}
    && S^{+}(x,x') \nonumber \\
    &=& \int d\mathbf{p} \int_{0}^{\infty} d\lambda \ \frac{|C_{\beta }^{(+)}|^2}{\omega} {(zz')}^{{D}/{2}}e^{i\mathbf{p(x-x')}-i\omega (t-t')} \ \     (\omega \gamma^0 - p_l \gamma^l + i\gamma^z \partial_z + \frac{i\gamma^z}{2z} + \frac{m}{kz})
       \nonumber  \\
     & &  \ \ \ \ \ \ \ \ \ \ \ \ \ \ \ \ \ \ \ \ \ \ \ \ \ \ \ \ \ \ \ \ \ \ \ \ \ \ \ \ \ \ \ \ \ \ \ \ \ \ \ \ \ \ \ \ \ \ \ \ \ \ \ \ \ \ \ \ \ \ \ \ \ \ \ \ \ \ \ \ 
     \times  \left[
     \begin{array}{cc}
         \widehat{J}_{+}(\lambda z)   \widehat{J}_{+}(\lambda z')  
    & \widehat{0} \\
      \widehat{0}   &
        \widehat{J}_{-}( \lambda z)  \widehat{J}_{-}( \lambda z') 
     \end{array}
     \right] \nonumber \\
     &=& \int d\mathbf{p} \int_{0}^{\infty} d\lambda \ \frac{|C_{\beta }^{(+)}|^2}{\omega} {(zz')}^{{1}/{2}} \ \     (i \gamma^0 \partial_0  + i  \gamma^l \partial_l + i\gamma^z \partial_z - \frac{i\gamma^z(D-1)}{2z}  + \frac{i\gamma^z}{2z} + \frac{m}{kz}) e^{i\mathbf{p(x-x')}-i\omega (t-t')}
       \nonumber  \\
     & &  \ \ \ \ \ \ \ \ \ \ \ \ \ \ \ \ \ \ \ \ \ \ \ \ \ \ \ \ \ \ \ \ \ \ \ \ \ \ \ \ \ \ \ \ \ \ \ \ \ \ \ \ \ \ \ \ \ \ \ \ \ \ \ \ \ \ 
     \times {(zz')}^{(D-1)/{2}}  \left[
     \begin{array}{cc}
         \widehat{J}_{+}(\lambda z)   \widehat{J}_{+}(\lambda z')  
    & \widehat{0} \\
      \widehat{0}   &
        \widehat{J}_{-}( \lambda z)  \widehat{J}_{-}( \lambda z') 
     \end{array}
     \right] \nonumber \\
     &=& \sqrt{\frac{z'}{z}} \left(i\left(\Gamma^{\mu}D_{\mu} + \frac{\Gamma^z}{2z}  \right) + m \right) \int \frac{d\mathbf{p}}{2\omega(2\pi)^{D-2}} \int_{0}^{\infty} d\lambda \ \lambda k^{D-1} e^{i\mathbf{p(x-x')}-i\omega (t-t')} {(zz')}^{(D-1)/{2}} \nonumber \\
     & &  \ \ \ \ \ \ \ \ \ \ \ \ \ \ \ \ \ \ \ \ \ \ \ \ \ \ \ \ \ \ \ \ \ \ \ \ \ \ \ \ \ \ \ \ \ \ \ \ \ \ \ \ \ \ \ \ \ \ \ \ \ \ \ \ \ \ \ \ \ \ \ \ \ \ \ \ \ \ \ \ \ \ \ \ \ \ \ \ \ \
     \times \left[
     \begin{array}{cc}
         \widehat{J}_{+}(\lambda z)   \widehat{J}_{+}(\lambda z')  
    & \widehat{0} \\
      \widehat{0}   &
        \widehat{J}_{-}( \lambda z)  \widehat{J}_{-}( \lambda z') 
     \end{array}
     \right] \nonumber \\
     &=& \sqrt{\frac{z'}{z}} \left(i\left(\Gamma^{\mu}D_{\mu} + \frac{\Gamma^z}{2z}  \right) + m \right) 
     \left( \mathcal{P}^{+}G_{AdS_D}(x,x',\frac{m}{k}+\frac{1}{2}) + \mathcal{P}^{-}G_{AdS_D}(x,x',\frac{m}{k}-\frac{1}{2})
     \right) \label{ssplus}
\end{eqnarray}
where $P^{\pm}= (\mathbb{I}_{N\times N} \pm i\gamma^z)/2$ and 
\begin{eqnarray}
    G_{AdS_D}(x,x',\frac{m}{k} \pm \frac{1}{2}) = \int \frac{{(zz')}^{(D-1)/{2}} d\mathbf{p}}{2\omega(2\pi)^{D-2}} \int_{0}^{\infty} d\lambda \ \lambda k^{D-1} e^{i\mathbf{p(x-x')}-i\omega (t-t')} J_{\frac{m}{k} \pm \frac{1}{2}}(\lambda z) J_{\frac{m}{k} \pm \frac{1}{2}}(\lambda z').
\end{eqnarray}

Just like equation (\ref{Splussss}) we can derive,
\begin{eqnarray}
    [S^{-}(x,x')]_{ab} &=& \bra{0} \overline{\Psi}_b(x') \Psi_a(x)\ket{0} \\ &=& \sum_{\sigma }  \int d\mathbf{p} \int_{0}^{\infty} d\lambda \  \psi^{(-)}_{\sigma}(\mathbf{p}, \lambda, x)_a    
    \overline{{\psi}_{\sigma}^{(-)}}(\mathbf{p}, \lambda, x')_b.
\end{eqnarray}
where $a,b$ are spinor indices. We can write this formula in matrix form,
\begin{eqnarray}
    S^{-}(x,x') = \sum_{\sigma }  \int d\mathbf{p} \int_{0}^{\infty} d\lambda \  \psi^{(-)}_{\sigma}(\mathbf{p}, \lambda, x)    
    \overline{{\psi}_{\sigma}^{(-)}}(\mathbf{p}, \lambda, x').
\end{eqnarray}
Next we can expand it using equation, (\ref{psi-})
\begin{eqnarray}
    & & S^{-}(x,x') \nonumber\\
    &=& \sum_{\sigma }  \int d\mathbf{p} \int_{0}^{\infty} d\lambda \ |C_{\beta }^{(-)}|^2 {(zz')}^{{D}/{2}}e^{i\mathbf{p(x-x')}+i\omega (t-t')}\left[
    \begin{array}{c}
    \frac{1}{\omega } \widehat{J}_{-}( \lambda z) \left[i\lambda
    -p_{l}\sigma^{l}\right) w^{(\sigma )} \\
    \widehat{J}_{+}( \lambda z) w^{(\sigma )}%
    \end{array} \right] \times \nonumber \\
    & &  \ \ \ \ \ \ \ \ \ \ \ \ \ \ \ \ \ \ \ \ \ \ \ \ \ \ \ \ \ \ \ \ \ \ \ \ \ \ \ \ \ \ \ \ \ \ \ \ \ \ \ \ \ \ \ \ \ \ \ \ \ \ \ \ \ \ \ \ \ \ \ \ \ \ \ \ \ \ \ \ 
    (i) \bigg[  w^{(\sigma ) \dagger} \widehat{J}_{+}( \lambda z') \ \ w^{(\sigma ) \dagger} \frac{\left(i\lambda
    +p_{q}\sigma^{q}\right)}{\omega} \widehat{J}_{+}(\lambda z')  \bigg] \nonumber\\
     &=& -\int d\mathbf{p} \int_{0}^{\infty} d\lambda \ |C_{\beta }^{(-)}|^2 {(zz')}^{{D}/{2}}e^{i\mathbf{p(x-x')}+i\omega (t-t')} (-i) \times \nonumber  \\
     & &  \left[
     \begin{array}{cc}
        \widehat{J}_{-}( \lambda z)  \frac{\left( i\lambda
    -p_{l}\sigma^{l}\right)}{\omega}\left( \sum_{\sigma } w^{(\sigma )}   w^{(\sigma ) \dagger} \right) \widehat{J}_{+}(\lambda z')  & \frac{1}{\omega }\widehat{J}_{-}( \lambda z)\left( i\lambda
    -p_{l}\sigma^{l}\right) \left( \sum_{\sigma } w^{(\sigma )}   w^{(\sigma ) \dagger} \right) \frac{\left( i\lambda
    +p_{q}\sigma^{q}\right)}{\omega} \widehat{J}_{+}( \lambda z')  \\
         \widehat{J}_{+}( \lambda z)\left( \sum_{\sigma } w^{(\sigma )}   w^{(\sigma ) \dagger} \right) \widehat{J}_{+}(\lambda z')  & \frac{1}{\omega }\widehat{J}_{+}( \lambda z) \left( \sum_{\sigma } w^{(\sigma )}   w^{(\sigma ) \dagger} \right) \left( i\lambda
    +p_{q}\sigma^{q}\right)  \widehat{J}_{-}( \lambda z') 
     \end{array}
     \right] \nonumber \\
      &=& -\int d\mathbf{p} \int_{0}^{\infty} d\lambda \ |C_{\beta }^{(-)}|^2 {(zz')}^{{D}/{2}}e^{i\mathbf{p(x-x')}+i\omega (t-t')} (-i) \times \nonumber  \\
     & &  \left[
     \begin{array}{cc}
       \frac{i\lambda}{\omega} \widehat{J}_{-}( \lambda z)\widehat{J}_{+}(\lambda z')  - \frac{\left(    p_{l}\sigma^{l}\right)}{\omega} \widehat{J}_{+}(\lambda z)   \widehat{J}_{+}(\lambda z')  
    &
    -\widehat{J}_{-}( \lambda z) \widehat{J}_{-}( \lambda z') \\
       \widehat{J}_{+}( \lambda z)  \widehat{J}_{+}(\lambda z')  
       &
       \frac{i\lambda}{\omega}\widehat{J}_{+}( \lambda z) \widehat{J}_{-}( \lambda z')  + \frac{\left( p_{l}\sigma^{l}\right)}{\omega} \widehat{J}_{-}( \lambda z)  \widehat{J}_{-}( \lambda z') \nonumber
     \end{array}
     \right] \nonumber \\
     &=& -\int d\mathbf{p} \int_{0}^{\infty} d\lambda \ \frac{|C_{\beta }^{(-)}|^2}{\omega} {(zz')}^{{D}/{2}}e^{i\mathbf{p(x-x')}+i\omega (t-t')} \ \  (-1) \times \nonumber  \\
     & & \left( i \left[
     \begin{array}{cc}
       {i\lambda} \widehat{J}_{-}( \lambda z)\widehat{J}_{+}(\lambda z') 
    & \widehat{0} \\
    \widehat{0} &
       i \lambda \widehat{J}_{+}( \lambda z) \widehat{J}_{-}( \lambda z')   
     \end{array} 
     \right] + (\omega \gamma^0 + p_l \gamma^l)
     \left[
     \begin{array}{cc}
         \widehat{J}_{+}(\lambda z)   \widehat{J}_{+}(\lambda z')  
    & \widehat{0} \\
      \widehat{0}   &
        \widehat{J}_{-}( \lambda z)  \widehat{J}_{-}( \lambda z') \nonumber
     \end{array}
     \right]
     \right) \nonumber\\
     &=& -\int d\mathbf{p} \int_{0}^{\infty} d\lambda \ \frac{|C_{\beta }^{(-)}|^2}{\omega} {(zz')}^{{D}/{2}}e^{i\mathbf{p(x-x')}+i\omega (t-t')} \ \     (-\omega \gamma^0 - p_l \gamma^l + i\gamma^z \partial_z + \frac{i\gamma^z}{2z} + \frac{m}{kz})
       \nonumber  \\
     & &  \ \ \ \ \ \ \ \ \ \ \ \ \ \ \ \ \ \ \ \ \ \ \ \ \ \ \ \ \ \ \ \ \ \ \ \ \ \ \ \ \ \ \ \ \ \ \ \ \ \ \ \ \ \ \ \ \ \ \ \ \ \ \ \ \ \ \ \ \ \ \ \ \ \ \ \ \ \ \ \ 
     \times  \left[
     \begin{array}{cc}
         \widehat{J}_{+}(\lambda z)   \widehat{J}_{+}(\lambda z')  
    & \widehat{0} \\
      \widehat{0}   &
        \widehat{J}_{-}( \lambda z)  \widehat{J}_{-}( \lambda z') 
     \end{array}
     \right] \nonumber\\
     &=&- \int d\mathbf{p} \int_{0}^{\infty} d\lambda \ \frac{|C_{\beta }^{(-)}|^2}{\omega} {(zz')}^{{1}/{2}} \ \     (i \gamma^0 \partial_0  + i  \gamma^l \partial_l + i\gamma^z \partial_z - \frac{i\gamma^z(D-1)}{2z}  + \frac{i\gamma^z}{2z} + \frac{m}{kz}) e^{i\mathbf{p(x-x')}+i\omega (t-t')}
       \nonumber  \\
     & &  \ \ \ \ \ \ \ \ \ \ \ \ \ \ \ \ \ \ \ \ \ \ \ \ \ \ \ \ \ \ \ \ \ \ \ \ \ \ \ \ \ \ \ \ \ \ \ \ \ \ \ \ \ \ \ \ \ \ \ \ \ \ \ \ \ \ 
     \times {(zz')}^{(D-1)/{2}}  \left[
     \begin{array}{cc}
         \widehat{J}_{+}(\lambda z)   \widehat{J}_{+}(\lambda z')  
    & \widehat{0} \\
      \widehat{0}   &
        \widehat{J}_{-}( \lambda z)  \widehat{J}_{-}( \lambda z') 
     \end{array}
     \right] \nonumber \\
     &=& - \sqrt{\frac{z'}{z}} \left(i\left(\Gamma^{\mu}D_{\mu} + \frac{\Gamma^z}{2z}  \right) + m \right) \int \frac{d\mathbf{p}}{2\omega(2\pi)^{D-2}} \int_{0}^{\infty} d\lambda \ \lambda k^{D-1} e^{i\mathbf{p(x'-x)}-i\omega (t'-t)} {(zz')}^{(D-1)/{2}} \nonumber \\
     & &  \ \ \ \ \ \ \ \ \ \ \ \ \ \ \ \ \ \ \ \ \ \ \ \ \ \ \ \ \ \ \ \ \ \ \ \ \ \ \ \ \ \ \ \ \ \ \ \ \ \ \ \ \ \ \ \ \ \ \ \ \ \ \ \ \ \ \ \ \ \ \ \ \ \ \ \ \ \ \ \ \ \ \ \ \ \ \ \ \ \
     \times \left[
     \begin{array}{cc}
         \widehat{J}_{+}(\lambda z)   \widehat{J}_{+}(\lambda z')  
    & \widehat{0} \\
      \widehat{0}   &
        \widehat{J}_{-}( \lambda z)  \widehat{J}_{-}( \lambda z') 
     \end{array}
     \right]  \label{psymmetry} \\
     &=& -\sqrt{\frac{z'}{z}} \left(i\left(\Gamma^{\mu}D_{\mu} + \frac{\Gamma^z}{2z}  \right) + m \right) 
     \left( \mathcal{P}^{+}G_{AdS_D}(x',x,\frac{m}{k}+\frac{1}{2}) + \mathcal{P}^{-}G_{AdS_D}(x',x,\frac{m}{k}-\frac{1}{2})
     \right) \label{sminus}.
\end{eqnarray}
Here in equation (\ref{psymmetry}) we have used the symmetry of integral that is $\mathbf{p} \to \mathbf{-p}$. Now we are going to consider the mass less case. We can take $m \to 0$ for equation (\ref{ssplus}) and (\ref{sminus}) for the massless limit. In order to do that we evaluate,
\begin{eqnarray}
    G_{AdS_D}(x,x',\pm \frac{1}{2}) &=& {(zz')}^{\frac{D-1}{2}} \int_{0}^{\infty} d\lambda \int \frac{e^{i\mathbf{p(x-x')}-i\omega (t-t')} d\mathbf{p}}{2\omega(2\pi)^{D-2}}  \ \lambda k^{D-1}  J_{ \pm \frac{1}{2}}(\lambda z) J_{ \pm \frac{1}{2}}(\lambda z') \nonumber \\
    &=& \frac{k^{D-2}{(zz')}^{\frac{D-1}{2}}}{2(2\pi)^{(D-1)/2}} \int_{0}^{\infty} d\lambda \  \left(\frac{\lambda}{\chi}\right)^{\frac{D-1}{2}}K_{\frac{D-3}{2}}(\lambda \chi)  \  J_{ \pm \frac{1}{2}}(\lambda z) J_{ \pm \frac{1}{2}}(\lambda z'), \label{GGads}
\end{eqnarray}
where $\chi = \sqrt{(\mathbf{x-x'})^2-(t-t'-i\varepsilon)^2}$ and $K_{\nu}(x)$ is modified Bessel function. Now we use  this relation in $J_{\pm}({\lambda z}) = \sqrt{2/(\pi \lambda z)} \sin(\lambda z + \pi /4 \mp \pi/4)$ in equation (\ref{GGads}).
\begin{eqnarray}
    & & G_{AdS_D}(x,x',\pm \frac{1}{2}) \nonumber \\
    &=& \frac{k^{D-2}{(zz')}^{\frac{D-3}{2}}}{(2\pi)^{(D+1)/2}} \int_{0}^{\infty} d\lambda \  \left(\frac{\lambda}{\chi}\right)^{\frac{D-3}{2}} K_{\frac{D-3}{2}}(\lambda \chi)  \ 2 \sin(\lambda z + \pi /4 \mp \pi/4) \sin(\lambda z' + \pi /4 \mp \pi/4) \nonumber\\
    &=& \frac{k^{D-2}{(zz')}^{\frac{D-3}{2}}}{(2\pi)^{(D+1)/2}}  \int_{0}^{\infty} d\lambda \  \left(\frac{\lambda}{\chi}\right)^{\frac{D-3}{2}}K_{\frac{D-3}{2}}(\lambda \chi) \left( cos(\lambda(z-z')) \pm cos(\lambda(z+z') ) \right). \label{Gpm}
\end{eqnarray}
The following integral\cite{prudinikov1992integrals} also helps us to simply the resutls,
\begin{eqnarray}
    \int_{0}^{\infty} d\lambda \  \lambda^{\frac{D-3}{2}}K_{\frac{D-3}{2}}(\lambda \chi)  cos(\lambda (z\pm z')) = \sqrt{\pi} \  2^{\frac{D-5}{2}}\chi^{\frac{D-3}{2}}\Gamma\left(\frac{D-2}{2}\right)((z\pm z')^2+\chi^2)^{\frac{2-D}{2}}. \label{refff}
\end{eqnarray}

Here $\Gamma(x)$ is gamma function. After some algebraic manipulation using equation (\ref{refff}), equation (\ref{Gpm}) becomes 
\begin{eqnarray}
    G_{AdS_D}(x,x',\pm \frac{1}{2}) = \frac{k^{D-2}\Gamma\left(\frac{D-2}{2}\right)}{2(2\pi)^{D/2}} \left((v-1)^{\frac{2-D}{2}} \mp (v+1)^{\frac{2-D}{2}} \right), \label{scalarGG}
\end{eqnarray}
where $v$ is conformal invariant distance defined as follows-\\
\begin{eqnarray}
    v = \frac{z^2+z'^2+(\mathbf{x-x'})^2-(t-t'-i\varepsilon)^2}{2zz'}. \label{numu}
\end{eqnarray}\\
From equation   (\ref{scalarGG})
and (\ref{numu})
, we can also note,
\begin{eqnarray}
    G(x,x',\pm \frac{1}{2}) &=& G(x',x,\pm \frac{1}{2}), \\
    S^{+}(x,x')_{m=0} &=& -S^{-}(x,x')_{m=0} .
\end{eqnarray}

\section{Acceleration in x-t path}
Here we show that the path in the $x^1-t$ plane considered in \eqref{eqnxt} actually results in uniform constant acceleration.
The components of the acceleration can be written as,
\begin{eqnarray}
    a^{(0)} &=& \frac{d^2 t}{d \tau^2} + 2 \Gamma^0_{0z} \left( \frac{dt}{d\tau} \right) \left( \frac{dz}{d\tau} \right) = z_0 k \omega \sinh{(\omega \tau)} \\ 
    a^{(1)} &=& \frac{d^2 x^{1}}{d \tau^2} + 2 \Gamma^1_{1z} \left( \frac{dx^{1}}{d\tau} \right) \left( \frac{dz}{d\tau} \right) = z_0 k \omega \cosh{(\omega \tau)}\\
    a^{(z)} &=& \frac{d^2 z}{d \tau^2} +  \Gamma^z_{zz} \left( \frac{dz}{d\tau} \right)^2 + \Gamma^z_{00} \left( \frac{dt}{d\tau} \right)^2+   \Gamma^z_{11} \left( \frac{dx^{1}}{d\tau} \right)^2 = -z_0 k^2 \\
    a^{(2)} &=&  a^{(3)} = \ ... \ = a^{(D-2)} = 0.
\end{eqnarray}

 Finally the magnitude of acceleration $\mathbf{a}$ becomes
\begin{eqnarray}
    |\mathbf{a}| = \sqrt{-a_{\mu}a^{\mu}} = \sqrt{-\left(g_{00}(a^{(0)})^2+g_{11}(a^{(1)})^2+g_{zz}(a^{(z)})^2 \right)} = a.
\end{eqnarray}
We have used the connection coefficients from appendix A. 
A point to note that we can not set $z_0=0$
to obtain linear uniformly accelerated 
path with finite acceleration.\\

\section{Conformal invariant for stationary trajectories}

A $D$ dimensional AdS spacetime $\mathcal{X}_{D}$ can be embedded in a $D+1$ dimensional flat spacetime $\mathcal{E}_{D+1}$ equipped with the metric $\eta_{\mu \nu}=diag(+1,-1,-1,\ldots ,-1,+1)$
in coordinates ${ (X_{0},X_{1},X_{2},\ldots,X_{D-1} ,X_{D})}$
with the following constrain,
\begin{eqnarray}
    F(X_0, \ldots , X_{D-1}) = \eta_{\mu \nu} X^{\mu} X^{\nu} - \frac{1}{k^2}=0. \label{eqconst}
\end{eqnarray}
Bros et al. \cite{bros} has demonstrated that the $2$ dimensional planer motion in $\mathcal{E}_{D+1}$ constrained to move in $\mathcal{X}_{D}$ would produce a constant acceleration in $\mathcal{X}_{D}$. The intersection between $\mathcal{X}_{D}$ and two dimensional  $\Pi$-plane could only produce parabola, ellipse or hyperbola (pair of straight lines as degenerate case). We are specially interested in hyperbolic paths over $\Pi$-plane. Due to Lorentzian structure of $\mathcal{X}_{D}$ it is possible to define time-like the tangent vector over $\mathcal{X}_{D}$ \cite{bros}. Our next discussion will closely follows the treatment give in \cite{bros}. The tangent vector $T^{\mu}$ over $\mathcal{X}_{D}$ is defined as time-like, light-like or space-like if $\eta_{\mu \nu}T^{\mu} T^{\nu}$ is greater than zero, equal to zero or less than zero respectively. 

Let $ X^{\mu}(\tau)$ denotes an arbitrary AdS trajectory parametrized by its proper time $\tau$ in $\mathcal{X}_{D}$. In general time-like  $V^\mu(\tau) = \frac{d} {d\tau} X^\mu(\tau)$ satisfies the following relations (in terms of scalar products in the ambient space $\mathcal{E}_{D}$): 
\begin{eqnarray}
    V^\mu(\tau)V_\mu(\tau) &=&  \eta_{\mu \nu} \frac{d} {d\tau} X^\mu(\tau) \frac{d} {d\tau} X^\nu(\tau)
     = \big( \frac{d \tau}{ d \tau} \big) ^2 = 1. \label{vvvmu} \\
     X^\mu(\tau) V_\mu(\tau) &=&  \eta_{\mu \nu}  X^\mu(\tau) \frac{d} {d\tau} X^\nu(\tau) = \frac{1}{2}   \frac{d }{ d \tau} \big( X^\mu(\tau) X_\mu(\tau )\big) = \frac{1}{2}  \big( \frac{d}{ d \tau} \frac{1}{k^2} \big) ^2 = 0. \label{xxvmu}
\end{eqnarray}
It follows that the ambient acceleration-vector $A^\mu(\tau) = \frac{d}{d\tau} V^\mu(\tau)$ satisfies the following relations.
\begin{eqnarray}
    X^\mu(\tau) A_\mu(\tau) & = & \eta_{\mu \nu}  X^\mu(\tau) \frac{d} {d\tau} V^\nu(\tau) = -\eta_{\mu \nu}  V^\mu(\tau)  \frac{d} {d\tau} X^\nu(\tau)  +  \frac{d} {d\tau} \big( X^\nu(\tau) V_\nu(\tau) \big) \nonumber \\
    & = & -V^\mu(\tau)V_\mu(\tau) +    \frac{d} {d\tau} (0) = -1. \label{axmu} \\
    V^\mu(\tau) A_\mu(\tau) & = & \eta_{\mu \nu}  V^\mu(\tau) \frac{d} {d\tau} V^\nu(\tau) = \frac{1}{2}   \frac{d }{ d \tau} \big( V^\mu(\tau) V_\mu(\tau )\big) = \frac{1}{2}   \frac{d}{ d \tau} (1)  = 0. \label{avmu}
\end{eqnarray}
Here we have used the results from equations (\ref{vvvmu}) and (\ref{xxvmu}) to derive the relation (\ref{axmu}) and (\ref{avmu}). If we want to derive acceleration vector $a^{\mu}(\tau)$ on $\mathcal{X}_{D}$ we need to take projection of $A^{\mu}(\tau)$ on the tangent hyperspace of $\mathcal{X}_{D}$ at point $P \in \mathcal{X}_{D}$. Let $X^{\mu}(\tau_0) = P^{\mu}$ be the position vector at point $P$ from center $O$. We will drop the $\tau$ dependence of the vectors when considering vectors at a  particular location in spacetime. Now the equation for tangent hyperspace would be 
\begin{eqnarray}
    (X^{\mu}-P^{\mu})\big(\nabla_{\mu}F(X_0, \ldots , X_{D-1})\big) \big|_{P} &=& 0 \nonumber \\ (X^{\mu}-P^{\mu}) \left(\frac{\partial}{\partial X^{\mu}}(X^{\nu}X_{\nu}-\frac{1}{k^2}) \right) \Big|_{P}  &=& 0 \nonumber \\
    k^2X^{\mu}P_{\mu} &=& 1. \label{tangenthyp}
\end{eqnarray}
This tangent hypersurface is clearly $D$ dimensional so we can project the vector $A^{\mu}$ on this hypersurface by subtracting a vector $B^{\mu}$. Now the vector $X^{\mu}
= P^{\mu}+a^{\mu} = P^{\mu} + A^{\mu} - B^{\mu}$ lies in the tangent hypersurface defined in equation (\ref{tangenthyp}) because $a^{\mu}$ has its base point at $P$. Therefore,
\begin{eqnarray}
    k^2(P^{\mu}+a^{\mu})P_{\mu} &=& 1 \nonumber \\ k^2\eta_{\mu \nu}(P^{\mu}+A^{\mu} - B^{\mu})P^{\nu} &=& 1 \nonumber \\
    \eta_{\mu \nu} B^{\mu}P_{\nu}P^{\nu} &=& \big(A^{\mu}X_{\mu} \big) \big|_{P} P_{\nu} \nonumber \\
    B^{\mu} &=& -{k^2}{P^{\mu}}.
\end{eqnarray}
Finally we can evaluate the magnitude of $a^{\mu}$ in terms of $A^{\mu}$.
\begin{eqnarray}
   a^{\mu}a_{\mu} &=& (A^{\mu}+k^2P^{\mu})(A_{\mu}+k^2P_{\mu}) \nonumber \\
   &=& \big( A^\mu  A_\mu  + 2 k^2 A^\mu  X_\mu + k^4 X^\mu  X_\mu \big) \big|_{P} \nonumber \\
   &=& A^\mu  A_\mu  - 2 k^2 + k^2 = A^\mu  A_\mu  -  k^2 . \label{eqAa}
\end{eqnarray}

Equation (\ref{eqAa}) implies that if the particle is moving in AdS spacetime and that particle's acceleration in flat spacetime $\mathcal{E}_{D+1}$ is constant then acceleration in $\mathcal{X}_{D}$ is also constant. In a similar fashion one can also extend the same conclusion for higher derivatives of the velocity vector. This conclusion could be used to classify the stationary motion in $D$ dimensional AdS spacetime from the classification of stationary motion in $D+1$ dimensional flat spacetime $\mathcal{E}_{D+1}$ by constraining the particle motion in $\mathcal{X}_{D}$. Here by stationary motion we meant the particle's world line is the orbit of some time-like Killing vector field. For stationary motions all Lorentz invariants constructed from the velocity and its derivatives must be constant \cite{Townsend}. Using this criterion it might be possible to classify all stationary paths in $\mathcal{X}_{D}$. In the later part of this appendix we outlined a formalism for constructing stationary trajectories which is closely related to Frenet-Serret formalism \cite{Iyer}.

\begin{figure}[t]
    \centering
    \includegraphics[scale=0.7]{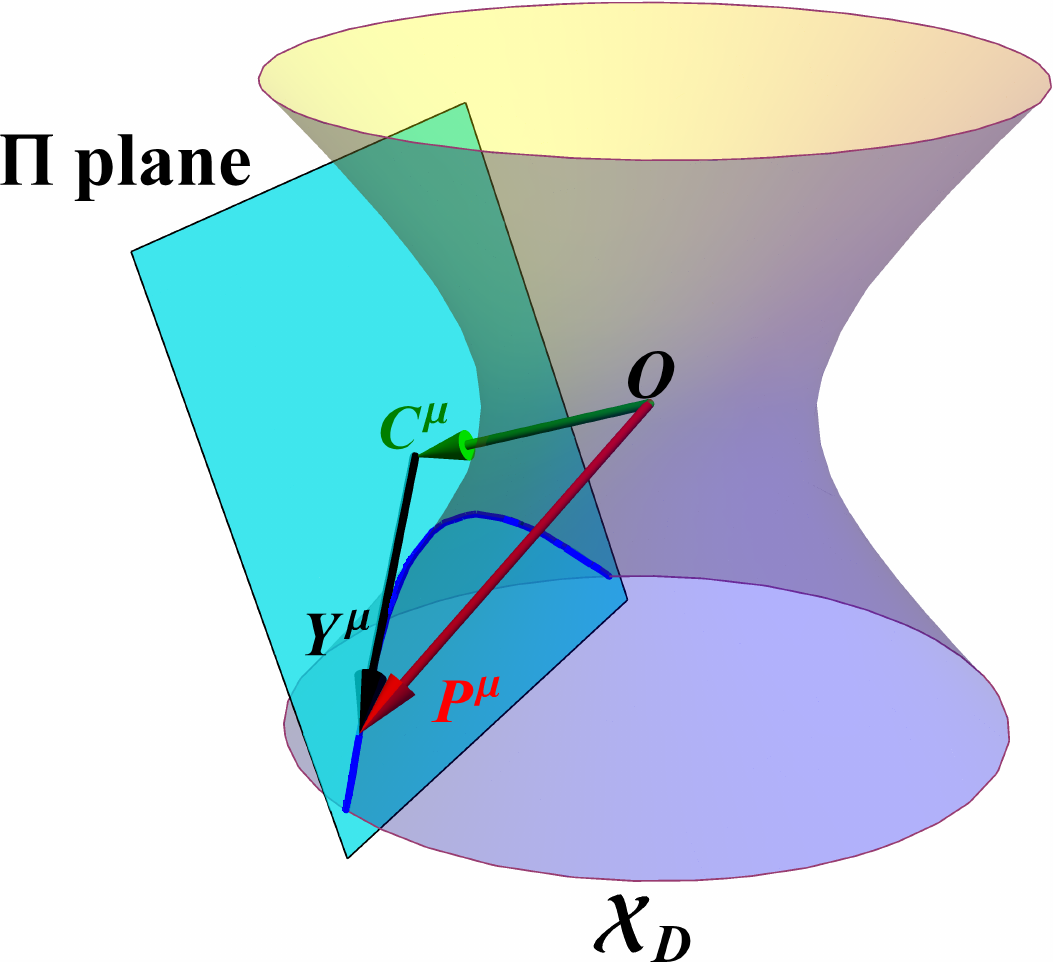}
    \caption{
    This figure depicts $D$ dimensional AdS hypersurface $\mathcal{X}_{D}$ embedded in $D+1$ dimensional flat spacetime $\mathcal{E}_{D+1}$. The blue curve slows the hyperbolic trajectory of a particle which is a conic section between two dimensional $\Pi$ plane and $\mathcal{X}_{D}$. Here $C^{\mu}$ is orthogonal to the $\Pi$ plane and $Y^{\mu}$ is contained in that plane. $P^{\mu}$ is a position vector of point $P$. The base point of this vector is the origin $O$ of $\mathcal{E}_{D+1}$.
    }
    \label{fig:fig6}
\end{figure}

In this paper our goal primarily is to work with particular class of stationary trajectory of particles. In this particular class, particle follows a path constructed from the intersection of a two dimensional plane $\Pi$ and $D$ dimensional AdS hypersurface $\mathcal{X}_D$. Now we are going to show that particle's acceleration on this type trajectory is constant and the motion is linear. Let's consider a particle's motion over the two dimensional $\Pi$ plane depicted in figure \ref{fig:fig6}. Here, $C^{\mu}$ is a vector perpendicular to $\Pi$ plane and a point $P$ lies on this plane too. Since the particle is moving on the two dimensional plane in flat spacetime $\mathcal{E}_{D+1}$, the velocity vector $V^{\mu}$ and $A^{\mu}$ must remain in this plane. Now we define a vector $Y^{\mu}(\tau) = X^{\mu}(\tau)-C^{\mu}$ that lies in $\Pi$ plane. Thus we can write
\begin{eqnarray}
    \frac{\partial Y^{\mu}(\tau)}{\partial \tau} \Bigg|_{P} = \frac{\partial (X^{\mu}(\tau)-C^{\mu})}{\partial \tau} \Bigg|_{P} = V^{\mu}(\tau) \big|_{P} = V^{\mu}, \label{yv}\\
     \frac{\partial^2 Y^{\mu}(\tau)}{\partial \tau^2} \Bigg|_{P} = \frac{\partial^2 (X^{\mu}(\tau)-C^{\mu})}{\partial \tau^2} \Bigg|_{P} = A^{\mu}(\tau) \big|_{P} = A^{\mu} \label{ya}.
\end{eqnarray}

From equation (\ref{yv}) and  (\ref{xxvmu}), it is easy to see that $V^{\mu}$ is orthogonal to $Y^{\mu}$ since $Y^{\mu}V_{\mu} = 0$. Similarly from equation (\ref{ya}) and (\ref{avmu}) we can deduce that $V^{\mu}$ is also orthogonal to $A^{\mu}$. As $A^{\mu}$, $V^{\mu}$ and $Y^{\mu}$ lie in a flat 2 dimensional plane, $A^{\mu}$ must be proportional to $Y^{\mu}$. We choose $A^{\mu} = \lambda Y^{\mu}$. We can determine the value of $\lambda$ in the following way. 
\begin{eqnarray}
    A^{\mu}(\tau)X_{\mu}(\tau)\big|_{P} &=& -1  \\
     \lambda Y^{\mu}P_{\mu} &=& -1 \label{eqlamyp} \\
     \lambda Y^{\mu}(P_{\mu}+C_{\mu}) &=& -1  \label{tricky}\\
     \lambda (P^{\mu} -C^{\mu}) (P_{\mu} +C_{\mu}) &=& -1 \\
     \lambda &=& \frac{k^2}{k^2 C^{\mu}C_{\mu}-1}.
\end{eqnarray}
In equation (\ref{tricky}) we have used the fact that $C^{\mu}$ is orthogonal to $Y^{\mu}$ because $Y^{\mu}$ lives in $\Pi$ plane. From here we can determine the magnitude of $A^{\mu}$.
\begin{eqnarray}
    A^{\mu}A_{\mu} &=& \lambda^2 Y^{\mu} Y_{\mu} \\
    &=&  \lambda Y^{\mu} (\lambda P_{\mu} - \lambda C_{\mu}) \\
    &=& -\lambda = \frac{k^2}{1-k^2 C^{\mu}C_{\mu}}. \label{aamu}
\end{eqnarray}
Here we used equation (\ref{eqlamyp}) in the last equation. We observe from the equation (\ref{aamu}) that the magnitude of vector $A^{\mu}$ doesn't depend on $P^{\mu}$. Therefore, the ambient acceleration has constant magnitude. Thus,
\begin{eqnarray}
    a^{\mu}(\tau)a_{\mu}(\tau) = A^{\mu}(\tau)A_{\mu}(\tau) - k^2 = \frac{k^4C^{\mu}C_{\mu}}{1-k^2 C^{\mu}C_{\mu}}. \label{constantproof}
\end{eqnarray}
Now lets put $a^{\mu}(\tau)a_{\mu}(\tau) = -a^2$ . This will give us $C^{\mu}C_{\mu} = (a^2 - k^2)^{-1} + k^{-2}$, $\lambda = a^2 - k^2 = \omega^2$ and $Y^{\mu}(\tau)Y_{\mu}(\tau) = - \omega^{-2}$. If we consider supercritical trajectory then $a > k$, thus $\lambda > 0$. From the above discussion we can write 
\begin{eqnarray}
    A(\tau) = \frac{\partial^2 Y(\tau)}{\partial \tau^2} = \omega^2 Y(\tau).
\end{eqnarray}
The solution for this differential equation is 
\begin{eqnarray}
    Y^{\mu}(\tau) = Y^{\mu}\cosh \big(\omega(\tau - \tau_0)\big) + \omega^{-1}V^{\mu} \sinh \big(\omega(\tau - \tau_0)\big) . \label{hyperbola}
\end{eqnarray} 
We used the position and velocity at point $P$ as initial condition e.g. $Y^{\mu}(\tau_0) = Y^{\mu}$ and $\dot{Y}^{\mu}(\tau_0) = V^{\mu}$. Here overdot represents derivative with respect to proper time. The equation (\ref{hyperbola}) represents hyperbola on $\Pi$ plane. We note that this hyperbolic trajectory can be easily interpreted as linear acceleration since over $\Pi$ plane $V^{\mu}$ and $Y^{\mu}$ are time-like and space-like vector respectively.  Finally we can evaluate the conformal invariant $v$ defined in equation ($\ref{nu}$) in Poincare coordinates. Poincare coordinates are defined by the following sets of equations,
\begin{eqnarray}
        X_0 & = &\frac{z}{2}\big( 1 + \frac{1/k^2 + |\vec{x}|^2 - t^2}{z^2} \big) \label{flat2pocare1}\\ 
        X_{D-1} & =& \frac{z}{2}\big( 1 - \frac{1/k^2 - |\vec{x}|^2 + t^2}{z^2} \big) \label{flat2pocare2}\\
    X_D & = &\frac{1}{k z}t \label{flat2pocare3}\\
    X_i & = &\frac{1}{k z}x_i. \label{flat2pocare4}
\end{eqnarray}
Here $i = 1, \ ... \ , D-2 $. Lets take two points on the trajectory at proper times $\tau =\tau_1$ and $\tau =\tau_2$. Using equations (\ref{flat2pocare1} - \ref{flat2pocare4}) the formula for conformal invariant $v$ can be written in the following form.
\begin{eqnarray}
    v(\tau_1,\tau_2) &=& k^2 X^{\mu}(\tau_1)X_{\mu}(\tau_2) \label{vformula} \\
    &=& k^2 \bigg( Y^{\mu}\cosh \big(\omega(\tau_1 - \tau_0)\big) + \omega^{-1}V^{\mu} \sinh \big(\omega(\tau_1 - \tau_0)\big) + C^{\mu} \bigg) \nonumber \\ &&\bigg( Y_{\mu}\cosh \big(\omega(\tau_2 - \tau_0)\big) + \omega^{-1}V_{\mu} \sinh \big(\omega(\tau_2 - \tau_0)\big) + C_{\mu} \bigg) \\
    &=& k^2 \bigg( Y^{\mu} Y_{\mu} \cosh \big(\omega(\tau_1 - \tau_0)\big)\cosh \big(\omega(\tau_2 - \tau_0)\big) + C^{\mu}C_{\mu} \nonumber \\
    && + \omega^{-2}V^{\mu}V_{\mu} \sinh \big(\omega(\tau_1 - \tau_0)\big)\sinh \big(\omega(\tau_2 - \tau_0)\big) \bigg) \\
    &=& k^2\bigg(-\omega^{-2} \cosh \big(\omega(\tau_1 - \tau_0)\big)\cosh \big(\omega(\tau_2 - \tau_0)\big) + \omega^{-2} + k^{-2} \nonumber \\
    & &-\omega^{-2}\sinh \big(\omega(\tau_1 - \tau_0)\big)\sinh \big(\omega(\tau_2 - \tau_0)\big)\bigg) \\
    &=& \frac{a^2}{\omega^2}-\frac{k^2}{\omega^2}\cosh \big(\omega(\tau_2 - \tau_1)\big). \label{numunumu}
\end{eqnarray}
In a similar manner one can derive the formula of $v(\tau_1, \tau_2)$ for subcritical paths in 2-plane (elliptic trajectory)  using $a< k, \ \lambda< 0$ \cite{Jennings}. Next we are going to illustrate two examples of uniformly accelerated supercritical trajectories. The supercritical trajectories for $z-t$ paths which are expressed as,
\cite{Jennings},
\begin{equation}
t(\tau)=\frac{a}{\omega}e^{\omega \tau} \;\;,\;\;
z(\tau)= e^{\omega\tau} \;\;,\;\;
x^{1} =x^{2}=x^{3}=\ldots=x^{D-2}=0.\nonumber
\end{equation}
In the higher dimensional flat spacetime $\mathcal{E}_{D+1}$, the  coordinates  take the following form on the path,
\begin{eqnarray}
       && X_0=\frac{1}{2}\big[z_0e^{\omega\tau}(1-\frac{a^2}{\omega^2})+
        e^{-\omega \tau}\frac{1}{k^2 z_0}  \big]\label{X_0},\\
         && X_{D-1}=\frac{1}{2}\big[z_0e^{\omega\tau}(1+\frac{a^2}{\omega^2})-
        e^{-\omega \tau}\frac{1}{k^2 z_0}  \big]\label{X_{D-1}},\\
&&        X_i=0,\label{qq}\\
&& X_D= \sqrt{\frac{1}{k^2}+\frac{1}{\omega^2}}.\label{kushas}
\end{eqnarray}
So. the $X_0$ and $X_{D-1}$ coordinates make the following relation,
\begin{eqnarray}
       X_{D-1}^2-X_0^2=\frac{1}{\omega^2}.\label{z-tpath}
\end{eqnarray}
Therefore eq. 
\eqref{qq}, \eqref{kushas}
and
\eqref{z-tpath}
parameterizes
  the super critical 
$z-t$ paths in $\mathcal{X}_{D}$ in the higher dimesnional flat spacetime $\mathcal{E}_{D+1}$.

Now the supercritical path in $x-t$
plane is given by,
\begin{eqnarray}
z(\tau)=z_0\;\;,\;\;
x^{1}(\tau)=\frac{z_0k}{\omega} \cosh(\omega \tau) \;\;,\;\;
t(\tau)=\frac{z_0k}{\omega}
\sinh(\omega \tau)
\;\;,\;\;
x^{2}=x^{3}=\ldots=x^{D-2}=0.\nonumber\label{55}
\end{eqnarray}
The higher dimensional flat space coordinates take the following form,
\begin{eqnarray}
&&X_0=\sqrt{\frac{1}{k^2}+\frac{1}{\omega^2}},\label{pixeqn1}\\
&&X_1=\frac{1}{\omega} \cosh(\omega \tau),\label{chandu1}\\
&&X_2=X_3=X_4=\ldots=X_{D-1}=0,\\
&&X_D=\frac{1}{\omega} \sinh(\omega\tau)\label{chandu2}.
\end{eqnarray}
We can clearly see from eq. \eqref{chandu1} and eq. \eqref{chandu2},\begin{eqnarray}
X_1^2-X_D^2=\frac{1}{\omega^2}. \label{ljylejhld}
\end{eqnarray}
So the $x-t$ path we described in \eqref{eqnxt} is for $D$ dimensional AdS spacetime.
In higher dimensional flat spacetime,
this path is actually the intersection of $\Pi_x$ two dimensional plane and $\mathcal{X}_{D}$. Here $\Pi_x$ plane  is defined as follows,
\begin{eqnarray}
\Pi_x \text{ plane:} \ X_{0} = \sqrt{\frac{1}{k^2}+\frac{1}{a^2-k^2}}, \  X_{2} = X_{3} = \ldots = X_{D-1} = 0.
\end{eqnarray}
The intersection is just an hyperbola $X_{1}^2-X_{D}^2 = \omega^{-2}$ hence the trajectory is supercritical \cite{bros}. However, 
 the $z-t$ 
path is actually the intersection of $\Pi_z$ plane and $AdS_{D}$ hypersurface, where $\Pi_z$ plane  is described by the following equations,
\begin{eqnarray}
    \Pi_z \text{ plane:} \ X_{D} = \sqrt{\frac{1}{k^2}+\frac{1}{a^2-k^2}}, \  X_{1} = X_{2} = \ldots = X_{D-2} = 0.
\end{eqnarray}
This plane can also be dubbed as $X_{D-1}-X_{0}$ plane where the hyperbola in equation \eqref{z-tpath} inhabits.
But, it is clearly apparent that $\Pi_x$ and $\Pi_z$ are the same plane with different parameterization over $D+1$ dimensional flat spacetime. If we rotate the coordinates in the following way-
\begin{eqnarray}
    &X_0& \leftrightarrow X_D \quad\text{and,} \\
    &X_1& \leftrightarrow X_{D-1},
\end{eqnarray}
the $\Pi_z$ plane becomes the $\Pi_x$ plane, the hyperbolic relation \eqref{z-tpath} turns to \eqref{ljylejhld} and vice versa. Therefore $z-t$  and $x-t$ paths are geometrically completely equivalent and they can be related to each other using AdS isometry.
We should point out if we took the $\Pi$-plane such as in the intersection we can have elliptic or parabolic equation in the intersection instead of hyperbolic such as \eqref{z-tpath}  or 
\eqref{ljylejhld}. However in such cases we will not have super critical paths in AdS. Instead we would have sub critical and critical acceleration in AdS for elliptic and parabolic paths in the intersection, respectively\cite{bros}.

We can extend this formalism to higher dimensions and generate stationary trajectories. The integral curve of time-like Killing vector vector field is defined as the stationary paths. Another equivalent description exists for the stationary trajectories \cite{Letaw}. If the geometric properties of any world lines are invariant under proper time translation, the path is called stationary path. For example the Whightman function is independent of proper time when the geodesic interval between two points only depends on the proper time interval between them \cite{Letaw}. For AdS spacetime if conformal invariant is a function of proper time interval then the world line is stationary. In order to investigate the geometric aspects of spacetime trajectories, the Frenet-Serret formalism is well suited \cite{Iyer}. Now we are going to outline our formalism which is closely related to the Frenet-Serret formalism. In the following discussion if a stationary trajectory is contained within the intersection of $M$-plane and $\mathcal{X}_{D}$, we will say the trajectory is contained in $M$ dimensions. This formalism can be used to categorize stationary trajectories in AdS by the minimum number of dimension  that would require to contain the path. 

In this formalism we are going to consider the world line in $\mathcal{E}_{D+1}$ confined within the intersection between $M (M<D+1)$ dimensional flat spacetime and $D$ dimensional $\mathcal{X}_{D}$ spacetime. Just like in figure (\ref{fig:fig6}) we define $C^{\mu}$ is perpendicular to the $M$-plane and point $P$ lies on the intersection. The vector $Y^{\mu}(\tau) = X^{\mu}(\tau) - C^{\mu}$ is contained with in the $M$-plane. Next we define the $n$th derivative of $Y^{\mu}(\tau)$ with respect to proper time 
\begin{eqnarray}
\frac{\partial^n Y^{\mu}(\tau)}{\partial \tau^n} = Y_{(n)}^{\mu}(\tau). \label{nderivative}
\end{eqnarray}

One important thing is to note here is that these vectors are defined in  $\mathcal{E}_{D+1}$. If we want to find these vectors defined over $\mathcal{X}_{D}$ we need to project them over the tangent hypersurface of $\mathcal{X}_{D}$ at some point $P \in \mathcal{X}_{D}$. We will use small case Latin letters to represent the projected vectors. After projection on the tangent hypersurface at point $P$ we find that 
\begin{eqnarray}
y^{\mu}_{(n)} = Y^{\mu}_{(n)} - k^2 ( Y^{\mu}_{(n)} Y_{(0){\mu}})\big|_P P^{\mu}.
\end{eqnarray}

These vectors are not necessarily orthonormal. We can construct a vielbein on the particle trajectory by using Gram-Schmidt orthogonalization process \cite{Letaw}. All possible inner products between two vectors will be used in the process. Moreover these inner products will also be used to construct the curvature invariants (curvature, torsion, hypertorsion etc.) of a particle's world line over $\mathcal{X}_{D}$. Frenet-Serret equations depend on these curvature invariants \cite{Letaw,Iyer}. If these curvature invariants are constant on the particle's trajectory, the particle's motion is stationary \cite{Letaw}. Therefore a sufficient condition to make a particle's world line stationary is that the inner products of all the pairs of $y^{\mu}_{(n)}$ are constant. The inner products 
\begin{eqnarray}
    y^{\mu}_{(n)}y_{(m)\mu} = Y^{\mu}_{(n)}Y_{(m)\mu} - k^2 (Y^{\mu}_{(n)}Y_{(0)\mu})\big|_P( Y^{\nu}_{(m)}Y_{(0)\nu})\big|_P \label{yYyY}
\end{eqnarray}
will be constant if and only if the inner products of all the pairs of  $Y^{\mu}_{(n)}$ are constant for all non-negative integer $n$. Now we need to find the inner products between all possible pairs of vectors defined in equation (\ref{nderivative}). We just state the results here. The derivation can be done using mathematical induction.
\begin{eqnarray}
Y^{\mu}_{(2p)}(\tau)Y_{(2q) \mu}(\tau) &=& (-1)^{p-q}Y^{\mu}_{(p+q)}(\tau)Y_{(p+q) \mu}(\tau) = (-1)^{p-q}Y^2_{(p+q)}(\tau). \label{inner1} \\
Y^{\mu}_{(2p+1)}(\tau)Y_{(2q+1) \mu}(\tau) &=& (-1)^{p-q}Y^{\mu}_{(p+q+1)}(\tau)Y_{(p+q+1) \mu}(\tau) = (-1)^{p-q}Y^2_{(p+q+1)}(\tau).  \\
Y^{\mu}_{(2p)}(\tau)Y_{(2q+1) \mu}(\tau) &=& 0. \label{inner2}
\end{eqnarray}
Here $p,q$ are non-negative integers. From the above discussion we can conclude if these inner products are constant, then the particle's world line in $\mathcal{X}_D$ is stationary. It also implies that these particles trajectories are also stationary in $\mathcal{E}_{D+1}$. However not all stationary world lines in $\mathcal{E}_{D+1}$ are stationary in $\mathcal{X}_{D}$ because those paths may not follow the constraint in equation (\ref{eqconst}). 
In order to construct stationary trajectories we need a proper notion of timelike direction which we already have defined using tangent vectors. Just like our previous discussion about the uniformly accelerated paths in 2-plane, we want to focus on the stationary paths in $\mathcal{X}_{D}$ that can only be contained within no less than $M$-plane. 

Lets assume $M$-plane can be spanned by $M$ number of linearly independent  $Y^{\mu}_{(0)},Y^{\mu}_{(1)}, \dots, Y^{\mu}_{(M-1)}$ vectors. These vectors are taken from the first $M-1$ derivatives of $Y^{\mu}_{(0)}(\tau = 0)$ at point $P$ on some stationary world line in $\mathcal{X}_{D}$. If these vectors are linearly dependent, we can confine the stationary trajectory within  dimensions less than $M$. Now we can construct an orthonormal basis of the  $M$-plane using these vectors using Gram-Schmidt orthogonalization process. If we choose appropriately the values of  $Y^{2}_{(p)}$ for $p>1$  and insists that $Y^{2}_{(1)} = 1, C^{\mu}C_{\mu} = |C|^2 > 1/k^2$, then the all members in the orthonormal basis except $Y^{\mu}_{(1)}$ can be made spacelike. This is always possible since inside the flat spacetime $\mathcal{E}_{D+1}$ any orthonormal basis could have at most two time-like vectors. In this situation the $M$-plane becomes Minkowski spacetime with only one time-like direction. However one can also work with other choices but the signature of embedding space $\mathcal{E}_{D+1}$ and constraint in equation (\ref{eqconst}) limit the choice of arbitrary values of these quantities.

Our goal now is to construct stationary paths that lies in the intersection between $\mathcal{X}_D$ and $M$-plane. We can employ Frenet-Serret formalism here to find the reduced linear differential equation for $Y^{\mu}_{(0)}(\tau)$ \cite{Letaw}. The key difference here from the usual Frenet-Serret formalism is that we can use $Y^{\mu}_{(0)}(\tau)$ due to the constraint in equation (\ref{eqconst}) and the solution would be a stationary path in $M$-plane just like the equation (\ref{hyperbola}) for 2-plane. If the world line can not be contained within no less than the $M$ dimension, the vielbein on the trajectory would have $M$ orthonormal vectors. The number of dimension wouldn't change with proper time, if the trajectory is stationary. Because all Lorentz invariants constructed from the velocity and its derivatives must be constant in this case \cite{Townsend}. It means that the inner products in equations (\ref{inner1}-\ref{inner2}) must be constant. Hence, the angle or relative location between $Y^{\mu}_{(0)}(\tau),Y^{\mu}_{(1)}(\tau), \dots, Y^{\mu}_{(M-1)}(\tau)$ will not change. So these basis vectors can be used to construct a vielbein. Thus $Y^{\mu}_{(M)}(\tau)$ can be expressed as linear combination of $Y^{\mu}_{(m)}(\tau)$ for $m=0,1, \ldots ,M-1$. We can express that in the following way. 
\begin{eqnarray}
Y^{\mu}_{(M)}(\tau) &=& \sum_{p=0}^{M-1} \lambda_p Y^{\mu}_{(p)}(\tau), \label{keyidea} \\
\frac{\partial^M Y^{\mu}_{(0)}(\tau)}{\partial \tau^M} &=& \sum_{p=0}^{M-1} \lambda_p \frac{\partial^p Y^{\mu}_{(0)}(\tau)}{\partial \tau^p}. \label{diffequationfrenet}
\end{eqnarray}
If the particle's motion is stationary, $\lambda_0, \lambda_1, \ldots , \lambda_{M-1} $ would be real numbers and they would not depend on proper time. We can uniquely determine these coefficients using the inner products defined in equations (\ref{inner1}-\ref{inner2}). We would have found the same equation (\ref{diffequationfrenet}), if we had used Frenet-Serret formalism to derive reduced linear differential equation for $Y^{\mu}_{(0)}(\tau)$. The solution of the differential equation would be
\begin{eqnarray}
Y^{\mu}_{(0)}(\tau) = \sum_{p=0}^{M-1}Y^{\mu}_{(p)} \ f_{p}(\tau, \lambda_0, \lambda_1, \ldots , \lambda_{M-1}). \label{solutions}
\end{eqnarray}
Here for $p=0,1, \ldots, M-1$, $f_{p}(\tau, \lambda_0, \lambda_1, \ldots , \lambda_{M-1})$ are some functions and $Y^{\mu}_{(p)}$ determined at the point $P$ are used as the initial conditions. Since $P \in M$-plane and the vielbein on each point of the trajectory has fixed $M$ number of basis vectors, that are just rotated form basis vectors of $M$-plane, the trajectory must contained within $M$-plane.  However, this solution doesn't necessarily confined in $\mathcal{X}_D$. Therefore we need to check whether $Y^{2}_{(0)}(\tau)$ is equal to $ 1/k^2-C^{\mu}C_{\mu}$ or not. If the solution satisfies the condition, then the conformal invariant $v(\tau_1,\tau_2)$ would be an even function of $\Delta \tau = \tau_1 - \tau_2$. From the equations (\ref{solutions}),(\ref{vformula}), (\ref{inner1}-\ref{inner2}) it is clear that $v(\tau_1,\tau_2)$ would only depend on $\Delta \tau, k^2, Y^2_{(0)}, Y^2_{(1)}, Y^2_{(2)}, \ldots, Y^2_{(M-1)}$.

In order to illustrate this formalism we select $M=2m+1<D+1$ dimensional plane which intersects $\mathcal{X}_{D}$. The values $Y^{2}_{(p)}$ for $p=0,1,2, \ldots, 2m$ are chosen appropriately so that the $M$-plane is a $2m+1$ dimensional Minkowski space. Now one can show that if $Y^2_{(m+p)} = (-1)^{p}\Omega^{2p}Y^2_{(m)}$ where $p \in \mathbb{N}$, the differential equation (\ref{diffequationfrenet}) becomes
\begin{eqnarray}
\frac{\partial^{2m+1} }{\partial \tau^{2m+1}}Y^{\mu}_{(0)}(\tau)  &=& \Omega^2 \frac{\partial^{2m-1} }{\partial \tau^{2m-1}} Y^{\mu}_{(0)}(\tau).
\end{eqnarray}
If $\Omega^2>0$, the solution of this linear differential equation is
\begin{eqnarray}
Y^{\mu}_{(0)}(\tau) &=& \sum_{q=0}^{2m-2}Y^{\mu}_{(q)}
\frac{\tau^q}{q!} + Y^{\mu}_{(2m-1)}
\left( \frac{\sinh(\Omega \tau)}{\Omega^{2m-1}} - \sum_{q=1}^{m-1} \frac{\tau^{2q-1}}{(2q-1)! \ \Omega^{2m-2q}}\right) \nonumber \\
&& +Y^{\mu}_{(2m)}
\left( \frac{\cosh(\Omega \tau)}{\Omega^{2m}} - \sum_{q=0}^{m-1} \frac{\tau^{2q}}{(2q)! \ \Omega^{2m-2q}}\right).
\end{eqnarray}
From this solution and using equation (\ref{vformula}) we can determine 
\begin{eqnarray}
 v(\tau_1, \tau_2) &=& k^2 \bigg( \frac{1}{k^2} - \frac{(-1)^{m} Y^2_{(m)}}{\Omega^{2m}} \Big(1 - \cosh \big(\Omega(\tau_1-\tau_2) \big) \Big) \nonumber \\
 &&+ \sum_{q=1}^{m-1}{\left(\frac{(-1)^q Y^2_{(q)}}{(2q)!}-\frac{(-1)^m Y^2_{(m)}}{(2q)!\Omega^{2m-2q}}\right)}(\tau_1-\tau_2)^{2q} \bigg).
\end{eqnarray}
From this example it is clear that $v$ is an even function of  
$\tau_1 - \tau_2$ and when $\tau_1 = \tau_2$, $v=1$.
Another example we want to show when a 4 dimensional plane intersects the $\mathcal{X}_D$ hypersurface. Then we can construct some new stationary trajectories which cannot be contained in dimension lower than 4. One such trajectory is as follows:
\begin{eqnarray}
&&X_0=|C|, \label{4planeeX1}\\
&&X_1 = \left(\frac{|Y|}{2} + \frac{1}{2|Y|\Omega^2} \right) \cosh(\Omega \tau) + \left(\frac{|Y|}{2} - \frac{1}{2|Y| \Omega^2} \right) \cos(\Omega \tau), \\ 
&&X_2 = \frac{1}{2\Omega^2}\sqrt{|A|^2-\frac{1}{|Y|^2}}\left(\cosh(\Omega \tau) - \cos(\Omega \tau)\right),\\
&&X_3 = \frac{\sqrt{|A|^4-\Omega^4}}{2\Omega^3}\left(\sinh(\Omega\tau)-\sin(\Omega\tau)\right),\\
&&X_4=X_5=X_6=\ldots=X_{D-1}=0,\\
&&X_D=\left(\frac{1}{2\Omega}+\frac{|A|}{2\Omega^3}\right)\sinh(\Omega \tau)+\left(\frac{1}{2\Omega}-\frac{|A|}{2\Omega^3}\right)\sin(\Omega \tau). \label{4planeeX2}
 \end{eqnarray}
In these equations $|C| = \sqrt{C^{\mu}C_{\mu}}>1/k$ and $|Y| =\sqrt{C^{\mu}C_{\mu}-1/k^2}= \sqrt{-Y_{(0)}^{\mu}Y_{(0)\mu}}$. The quantity $|A| = \sqrt{-Y_{(2)}^{\mu}Y_{(2)\mu}} = \omega = \sqrt{a^2 - k^2} > 0$ is the magnitude of acceleration in $\mathcal{E}_{D+1}$. The absolute value of the magnitude of second derivative of velocity is $|Y_{(3)}|=\sqrt{-Y_{(3)}^{\mu}Y_{(3)\mu}} = \Omega^2 = |A|/|Y|$. If we apply this condition to the equation (\ref{diffequationfrenet}), we will obtain  
\begin{eqnarray}
\frac{\partial^4 Y^{\mu}_{(0)}(\tau)}{\partial \tau^4} = \Omega^4 Y^{\mu}_{(0)}(\tau). \label{hudai}
\end{eqnarray}
In general for any 4-plane intersection with $\mathcal{X}_{D}$ with the above conditions one can derive the conformal invariant 
\begin{eqnarray}
v(\tau_1, \tau_2) = k^2 \left(|C|^2 + \left(\frac{|Y|^2}{2} + \frac{1}{2\Omega^2} \right) \cosh(\Omega (\tau_1-\tau_2)) + \left(\frac{|Y|^2}{2} - \frac{1}{2\Omega^2} \right) \cos(\Omega (\tau_1-\tau_2))\right). \label{nulast}
\end{eqnarray}
In the computation of 4 dimensional stationary path we could independently choose the values of $Y_{(0)}^2$ and $Y_{(2)}^2$ provided $|Y_{(2)}^2| > 1/|Y_{(0)}^2|$. The magnitude of all other higher derivatives of $Y^{\mu}_{(0)}(\tau)$ can be determined from these two values using the relation (\ref{hudai}). However if we put $|Y_{(2)}^2| = 1/|Y_{(0)}^2|$, the trajectory defined in equations (\ref{4planeeX1}-\ref{4planeeX2}) will reduce to the uniformly linear accelerated path defined in the  equations (\ref{pixeqn1}-\ref{chandu2}). Similarly the formula of $v(\tau_1,\tau_2)$ in equation (\ref{nulast}) will change to (\ref{numunumu}). 

This example suggests that the magnitude of $Y^{\mu}_{(0)}(\tau)$ and higher derivatives of $Y^{\mu}_{(0)}(\tau)$ determine the minimum number of dimensions one must need to contain any stationary trajectory in $\mathcal{X}_{D}$. Given all the values of $Y^{2}_{(p)}$ $(p \geq 0)$, we can find the number of orthonormal basis vectors using Gram-Schmidt orthogonalization process. It is always possible to do so, because all possible inner products in equations (\ref{inner1}-\ref{inner2}) depend on these values. For example if we have $Y^2_{(r)} = (-\Omega^2)^{r-1}$ for non-negative integer $r$, it is easy to show that only two vectors would survive the Gram-Schmidt orthogonalization. Hence, with these values of $Y^2_{(r)}$ the particle's world line must be confined within 2-plane and the particle is uniformly accelerating. Furthermore, if we already know the world line is contained within at least $N$ dimension, then we would only require squared norm of the first $N$  derivatives of $Y^{\mu}_{(0)}$ to figure out the minimum number of dimension required to contain that path.  Therefore, in $\mathcal{X}_D$ the minimum number of dimensions would require to contain a stationary world lines and the conformal invariant $v(\tau_1, \tau_2)$ of the path can uniquely be determined from the values of $Y^{2}_{(p)}$  ($0 \leq p \leq D$).

Now lets return to the local description of stationary paths in $\mathcal{X}_D$ and lets suppose that we are given all the inner products of the form $y^{\mu}_{(q)}y_{(q)\mu}$ for $2 \leq q \leq D+1$. From the equation (\ref{yYyY}) using mathematical induction it is possible to reproduce  all the values of $Y^{2}_{(q)}$. From these values we can actually find the minimal confinement dimension of the stationary trajectory inside $\mathcal{E}_{D+1}$. After that we can follow our formalism described earlier and work our way upto equation (\ref{keyidea}). Next we differentiate both sides with respect to proper time and then use $Y^{2}_{(q)}$s to find the values of $\lambda_q$s. Using these values it is easy to extract the value of $Y^{2}_{(0)}$ from the equation (\ref{keyidea}). So we found all the values of $Y^{2}_{(p)}$ $(0 \leq p \leq D)$. From this we can conclude that in order to uniquely determine the $v(\tau_1, \tau_2)$ of any stationary path in $D$ dimensional AdS spacetime, we only need the squared norm of first $D$ number of covariant derivatives of velocity vector with respect to propertime (  $y^{\mu}_{(q)}y_{(q)\mu}$ , $q =2, \ldots, D+1$).
\bibliographystyle{unsrt}  


\begin{thebibliography}{1}
\bibitem{y}We specially thank the referee to bring our attention to this problem. We have extended our analysis in Appendix $C$ in the revised version and gave focus on a new formalism, much like Frenet-Serret formalism, 
 that was used to construct and categorize stationary trajectories in $AdS_D$ spacetime based on
 Referee's comments.
\end{thebibliography}

\end{document}